# AI Alignment as a Systems Problem: Toward Distributed Proof Checking at Runtime

B. Gopinath[1]

Systems AI Labs LLC

## Abstract

AI alignment concerns the behavior of a composed system: what it must be able to do as well as what it is permitted to do. This paper returns to IC*, a language-independent model of computation in which invariants govern admissible state changes. Its development into agreement-controlled, nested logical times supplies the systems foundation. We give a relational construction in which each component has a state invariant, transition law, vote, and moves between ticks. A parent clock checks its children's agreement and joint-transition predicates; separate child ticks require parent permission. The resulting clock is itself an object, allowing arbitrary nesting.

The main theorem derives global refinement from local obligations: faithful, refining leaves and sound, refining connecting clocks yield a faithful hierarchy refining its root specification. A traditional proof accompanies the Lean formalization. A matching proposal-and-commit protocol is sound and, under finite branching, complete for enabled ticks. Finite trace coverage identifies the additional witness obligation needed to retain required capabilities. The development checks cause-and-effect obligations at their governing clocks and applies the construction to supervised AI teams. Justified local transitions compose into histories satisfying specified predicates.

The paper proposes distributed proof checking using Lean instances connected through verified logical-clock protocols. A Lean-checked evolving-forest preservation theorem gives conditions for authorized creation, alongside a traditional induction proof. Retaining commitments, logical epochs, and supporting evidence could provide a verifiable ledger. Realizing the architecture requires proving correspondence between the distributed implementation, its external effects, and the model.

## 1 Introduction and contribution

Artificial intelligence has entered a regime in which the model is no longer always the natural boundary of the system being engineered. Foundation models are embedded in products, toolchains, organizational workflows, autonomous agents, vehicles, communication systems, and multi-agent environments. They receive observations from changing environments, act on those environments, influence human behavior, generate new data, and are subsequently updated in response to the consequences of earlier deployment. In such settings, the object that must remain safe and useful is not merely a parameterized model. It is a closed loop involving the model, its users, other software and agents, institutional rules, physical or digital environments, and the feedback pathways that connect them.

The specific problem studied here is how to verify a hierarchy whose components can advance at different logical rates. Joint steps must satisfy shared constraints, while locally permitted steps must preserve the obligations of the enclosing clocks. Can a parent's obligations be discharged over the specifications of its immediate children, with an explicit account of autonomy and time?

[1] Correspondence: gopi@gopi.ai

Training language models to follow human preferences is a substantial achievement [1, 2]. Yet the same model can produce different consequences when its permissions, memory, tools, users, or institutional setting change. A response that is harmless as advice may become consequential when it directly initiates a payment, modifies a database, or delegates authority. Evaluating the model alone cannot establish the behavior of all such compositions.

Existing safety research extends beyond training. Shielding, verification of learning-enabled control, runtime assurance, and safety cases address systems in operational environments [3–6]. Systemic alignment is proposed here as an organizing perspective: the evolving closed loop is the primary object of assurance, and behavioral guarantees depend on the continuing validity of their assumptions.

Systemic alignment is defined here as the maintenance of an AI-enabled sociotechnical system within explicitly defined acceptable behavioral bounds as its components, users, institutions, and environment change. An acceptable specification may include safety, task performance, privacy, availability, and accountability. A formal certificate establishes only the properties actually expressed in that specification, under its stated assumptions. Selecting legitimate bounds remains a human and institutional responsibility.

The cybernetic connection is observation, intervention, and feedback. A system must obtain enough information about the relevant state to justify its actions, and those actions must preserve the chosen properties under admitted uncertainty. In the present model, the obligations governing a proposed transition are satisfied before its logical clock advances. That transition may create obligations for later cycles. Section 4.4 specifies observation at a clock's own time, Section 7 treats cause and effect obligations, and Section 8 addresses the conditions under which the model applies.

Building on IC*/CO*'s invariant-centered model and agreement-controlled logical time, this paper gives a relational construction and machine-checked compositional verification rule. Its invariant, transition-law, vote, and inner-move interface makes a connecting clock an object for its parent. Local obligations yield global refinement while preserving logical tick counts, including separate child advancement under parent permission. A matching tentative-state protocol is sound and, under finite branching, complete for enabled ticks; finite trace coverage identifies the additional witnesses required for useful capabilities. The evolving-namespace extension checks preservation under authorized creation. The AI-team construction applies the interface to guarded commitments and delegated authority. Communicating Lean instances and their implementation correspondence remain the proposed application in Section 10.

Refinement restricts implementation behavior to the specification. Protocol completeness preserves the ability to realize semantically enabled transitions. Required behavior must additionally have implementation witnesses; Section 3 states the two behavioral bounds and Section 5.3 gives their finite trace correspondence. Progress obligations, including fairness and timely service, are collected in Section 12.1.

The construction has a published lineage. IC* defined systems through invariants and constrained state changes in 1988 [7–9]. Das, Gopinath, and D. Kurshan's 1991 CO* account connected this approach to agreement-controlled logical times, nested finer times, and histories represented by forests with equivalences [10]. The present paper formalizes a selected compositional core, proves its refinement and protocol results, and examines its use for AI alignment. L.0 is an executable language instance in this lineage, rather than the origin of the model.

The body explains the definitions and proofs in English. References such as A4.9 identify exact statements in the appendix, whose original numbering is retained. Section 2 introduces the IC*/CO* foundation and its systems background. Sections 3–8 develop the specifications, construction, proof, language instance, and applicability obligations. Section 9 treats supervised AI teams. Section 10 proposes distributed proof checking and optional history retention; Sections 11 and 12 locate the contribution and state its limits and reproducibility evidence.

## 2 IC* foundations and systems background

My engagement with these questions began at Stanford in the 1960s and continued at Bell Labs. Being taught by Paul Cohen (winner of the Fields Medal), Rudolf Kalman (winner of the Kyoto Prize), Thomas Kailath, and Benjamin Otto Lange left one lasting impression: that applied mathematics must first be mathematics. The practical reach of observer and controller mathematics—from spacecraft to everyday machines—reinforced that conviction. A mathematical formulation can travel across domains when its assumptions, conclusions, and limits are precise and its construction can be reliably reproduced. This paper continues that arc: the search for a formulation of AI alignment that supports both mathematical verification and practical reuse. The episodes below explain how these questions recurred in my work, with published results cited where available.

### The IC* invariant model

IC* is a model of computation that can support more than one programming language [7–9]. Its 1988 account defines state and activation histories, with static, differential, and conditional invariants governing admissible behavior. Static invariants constrain states. Differential invariants constrain atomic changes of the relevant substate and may refer to earlier values. Conditional invariants control the activation and termination of obligations, including historically bound values. The state-construction procedure seeks values satisfying the applicable constraints and rejects a proposed construction that violates them.

This is the sense in which I call the model self-proving, or designed for provability. Its specification supplies predicates that the modeled execution must maintain; a valid initial state and justified transitions support an inductive proof of the resulting history. The present paper makes the local preservation and composition obligations explicit and checks their consequences in Lean. Establishing that an implementation faithfully enforces those obligations is the separate correspondence task in Section 8.2.

The published IC* semantics indexes a history by discrete steps. Such an index orders modeled atomic events; it need not be an explicit physical clock used to coordinate execution. The step boundaries remain part of the semantics: refining a time representation must preserve the events and historical bindings on which the predicates depend. Here the chosen boundaries are specified by each object's ticks and by the parent's rules for joint and separate advancement.

### CO* and agreement controlled logical time

Das, Gopinath, and D. Kurshan's Foundations of concurrency among objects [10] explicitly extends IC* and L.0 through the CO* model. It describes logical times whose advancement is subject to participating objects' agreement, finer private times nested beneath shared coarser times, and object histories represented by evolving forests with equivalences. Thus the logical-time hierarchy is part of the published development of the model. The paper also connects the model to self-timed computation and to algorithms for retrieving structural history.

The related protection, coordination, and persistence work describes interface cycles that advance only when their conditions are discharged, while nested cycles can defer an enclosing advance [11]. The parent-clock construction used here makes that discipline explicit: the parent checks each immediate child's vote and its own joint predicates. Checking can be centralized at that parent, but a joint tick still requires every child's specified agreement. At the root, the protocol defines the final commit directly.

### State Observers and the Autopilot Problem

An early motivation came from an autopilot problem brought to my attention by my doctoral advisor. In my recollection, test pilots were uncomfortable with aircraft in which the autopilot had separated the joystick from the control surfaces. The haptic signals through which an experienced pilot sensed the aircraft's behavior were missing. This was an observation problem as well as a control problem: precise automatic control did not by itself give the pilot the information needed to judge the aircraft's condition.

I proposed driving the joystick from an estimate of the aircraft's state while retaining automatic control of the surfaces, with a switch to return control authority to the pilot. The conceptual separation was useful. Measurable outputs could drive a parallel dynamical system that reconstructed relevant state, and that estimate could support either automatic control or the operator's awareness.

My doctoral work formalized observer and controller constructions for linear systems [12, 13]. In the ideal observable linear model, observer error dynamics can be assigned through suitable pole placement. That statement does not imply arbitrarily precise or arbitrarily fast estimation in a physical implementation. Measurement noise, model error, numerical conditioning, and bandwidth constrain performance. Subsequent induction-machine flux-observer research illustrates the practical relevance of the observer approach [14].

The lasting lesson is that direct measurement of every internal variable is unnecessary when a model and available measurements support a reliable estimate. Observability and controllability are dual in classical linear theory [15]. For alignment, the relevant question is whether the system can estimate enough of the state, with adequate uncertainty bounds, to support its required interventions. This motivates the clock-relative observation obligations in Section 4.4.

### Queuing and Deadlock Freedom

At Bell Labs, data networks brought shared buffers, stochastic arrivals, and resource dependencies into the foreground. We used symbolic computation to derive analytical expressions for queuing models. Buffer allocation raised the separate question of excluding cycles of mutual waiting. These experiences distinguished a restriction on behavior from a guarantee of progress: an invariant can exclude a forbidden configuration while fairness and service remain additional obligations. That distinction reappears in Sections 3 and 12.1.

### Synchronizing the Switching Network

Digital switches rely on physical oscillators. Differences in local frequency cause incoming and outgoing streams to drift, producing dropped or repeated frames. Pierce described the network synchronization problem [16]. The apparently digital semantics of a connection therefore depended on the behavior of an analog substrate.

In my work with Debasis Mitra and Irwin Sandberg, we studied coupled differential equations and used Lyapunov reasoning to analyze clock behavior. Feedback used measured phase differences to adjust local frequency. Under the hypotheses of the synchronization model, a suitable Lyapunov argument established convergence or bounded error. Related published analysis of a hybrid master-slave and mutual synchronization architecture appears in Mitra [17].

Delay, topology, coupling laws, and implementation accuracy determine whether that conditional guarantee applies to an operating network. This is the antecedent of the applicability and monitoring obligations in Section 8: the closed loop must remain within the conditions under which its proof is valid.

### Periodic Generators and Protocol Verification

Kurshan and I collaborated on digital signaling, including Touch-Tone receiver-generators and single-tone generator-detectors [18, 19]. Cyclotomic recursions reduced the need for expensive multiplication. Our 1974 result concerns a generator analytic near a fixed point that is periodic as a map: a finite iterate is the identity. Under those hypotheses, its Jacobian at the fixed point has exactly the same finite order [20]. The hypothesis concerns the map, rather than one periodic orbit. The lesson carried forward here is to identify the mathematical class for which a certificate controls the claimed behavior.

In protocol design, Kurshan and I also used a lower language of required behaviors and an upper language of permitted behaviors to bound a system language. Both bounds were representable by finite-state machines in a key case. This account of the earlier two-bound formulation is my historical recollection; the published S/R formalism and COSPAN use finite-state concurrent processes and verification through language containment [21]. Section 3 retains both bounds, and Section 5.3 separates the protocol's preservation of enabled behavior from the design obligation to supply useful witnesses.

#### L.0 as an executable language instance

L.0 provides a language and environment for specifying, simulating, and prototyping concurrent protocols within this model lineage [22, 23]. Linda Ness supplied its formal semantic account and analyzed associated verification issues [24]. The work was developed in the setting of protocol and service creation; Ness's account uses Futurebus+ as a working example. Here L.0 supplies a concrete cause and effect core through which the logical-clock construction can be examined in Section 7.

Lean [25] now permits these selected obligations to be stated and checked explicitly. The checked development in this paper covers the object and clock interfaces, composition, a matching protocol, and a core of L.0 formulas. Its scope is stated separately from the full IC* model, the CO* history representation, and the complete L.0 language.

#### Histories and Temporal Databases

With Vassilis Tsotras and colleagues, I worked on efficient management of evolving databases, forests, object versions, and temporal views [26–30]. Optimally managing the history of an evolving forest [27] addresses the representation and retrieval of structural history. The 1991 CO* paper explicitly connects its evolving forest representation to this work [10]. Section 10.3 builds on that connection: retained commitments would additionally carry their logical epochs and the evidence authorizing them.

A past fact is interpreted in its historical context. In the proposed architecture, an occurrence is identified by its component and logical epoch, whose number records advancement under governing predicates. Retaining the referenced state and proof context supports later verification. Efficient history management makes the relevant past accessible; the deployment obligations in Section 8.2 establish how recorded commitments correspond to external events.

IC* supplies the invariant-centered model; CO* supplies its published logical-time and structural-history development; L.0 supplies an executable language instance. The earlier systems work explains why observation, admissible behavior, and the continued applicability of a model belong in the same account. The following sections give the selected compositional construction its definitions and proofs.

## 3 Required and permitted behavioral languages

### 3.1 Both bounds are obligations

Let $L_{system}$ be the set of behaviors admitted by a closed-loop model. Let $L_{min}$ specify behaviors that the system must be capable of realizing, and let $L_{max}$ specify the behaviors it is permitted to produce. The full behavioral-bounding obligation is

$$L_{min} \subseteq L_{system} \subseteq L_{max}.$$

The lower inclusion requires behavioral coverage: every trace in the lower language must be realizable by the modeled system. The upper inclusion restricts behavior: every trace the system can produce must lie within the permitted language. A necessary consistency condition is $L_{min} \subseteq L_{max}$; a behavior cannot be required while also excluded. This condition alone does not establish that an implementation can meet both bounds under its environmental assumptions.

For each application, the alphabet, trace interpretation, system boundary, environment, and specification must be defined. Finite traces express many safety requirements; infinite traces with suitable acceptance conditions can express progress. The languages need not be regular, and general containment need not be decidable. An application must justify the abstraction and establish the bounds it claims.

An upper bound alone admits a system that refuses every useful task. A lower language containing required completed behaviors excludes universal refusal by requiring those behaviors to remain realizable. Both language bounds therefore belong in the specification. Lower inclusion is existential coverage of traces: stronger guarantees of response, availability, fairness, and eventual completion require progress conditions and environmental assumptions.

The lower language must distinguish system choices from events the system cannot ensure. A required behavior may depend on an available tool, an authorized request, or a cooperative environment. Those conditions belong in its interpretation. A universal guarantee and a probabilistic guarantee are also different claims; the latter requires a stated probability model and risk threshold. Neither should be inferred from an unqualified containment formula.

### 3.2 The language established by the formal development

The appendix's SafeTrace and Certificate framework defines upper containment (A2.6–A2.12). Faithfulness and refinement later supply preservation and simulation obligations. They do not by themselves establish the lower inclusion for an application's chosen required language. Protocol completeness concerns the concrete object's own enabled votes. It does not make every abstractly permitted transition implementable.

Section 5 states a finite coverage corollary for root-only commits. It identifies a precise language for which soundness and completeness supply both directions of correspondence. To obtain an application's lower bound, one must additionally show that every required observable trace has a concrete semantic witness. This separates preservation of capabilities by the protocol from the substantive question of which useful capabilities the design actually provides.

### 3.3 Bounds across authorized change

An evolving system needs a versioned interpretation of both bounds. Let k identify a coherent version of the implementation model, assumptions, lower and upper specifications, and certificate. At decision time t, let H be the recorded history. Define the system language below as the continuations feasible from that history under the assumptions for version k. The lower language specifies required continuations that must remain realizable, and the upper language specifies permitted continuations when appended to that history. The local obligation is

$$L_{min}^{k}\left(H_t\right) \subseteq L_{system}^{k}\left(H_t \vee A^k\right) \subseteq L_{max}^{k}\left(H_t\right).$$

Changing a model or policy must preserve the interpretation of pending actions and earlier commitments. A revised bound cannot make an earlier violation disappear. A transition between versions requires compatible state, jointly satisfiable assumptions, preservation of required capabilities and continuing obligations, and installation of the matching checks and enforcement rules. A record of the change identifies the version; it does not prove that the change preserves either inclusion.

# 4 Objects, nested clocks, and logical time

## 4.1 Object and clock interfaces

An *object* is the unit of the theory. It has a state. It has a *static invariant*, a condition its state must always satisfy. It has a *law of the tick*, a condition relating its state before a tick to its state after. It has a *vote*: for each proposed change of its state, it either agrees or does not. And it has *inner moves*, changes it makes by itself between ticks of the clock it is attached to (A4.1).

An object is *faithful* when it keeps its word. It votes only for changes that obey its law and preserve its static invariant, and its inner moves preserve its static invariant (A4.2).

Logical time counts a clock's ticks. A tick is an agreed transition at that clock; a commit is the protocol operation that realizes it. An epoch is the interval of a clock's history ending at one of its ticks, including any permitted intervening activity. Parent and child clocks count their own ticks, so their epoch numbers need not be equal. Historical bindings identify the component and epoch to which a fact belongs. A physical interpretation may associate epochs with intervals of real time; the abstract counter itself measures events, not duration.

A clock carries a family of child objects, an own predicate on their joint state, a predicate on the joint before-and-after state, and an allow relation for a child's separate tick (A4.5). A joint tick is licensed exactly when every child votes for its component of the proposed change and the joint transition predicate holds. This is a transition relation, not a scheduling rule: an enabled tick is not forced to happen. Failure to obtain the joint evidence prevents that proposed joint commit; separately permitted child activity may still change the joint state.

The clock constructor turns the family into an object (A4.7). Its static invariant combines the children's invariants with the clock's own predicate. Its tick law combines their laws with the joint transition predicate. Its inner moves consist of an allowed separate tick of one child or an inner move within one child. Because the result has the same object interface, it can be placed under another clock.

A sound clock preserves its own predicate under both its licensed joint ticks and its inner moves, from states satisfying the composite invariant (A4.8). This is an obligation to prove, not a consequence of declaring a guardrail. Closure follows: a sound clock with faithful children is a faithful object (A4.9). In a joint tick, each faithful child preserves its own invariant and obeys its law; soundness preserves the parent's predicate. The corresponding argument covers inner moves. Repeating the argument up the tree preserves the composite invariant (A4.10, A4.16).

## 4.2 Joint ticks and autonomous child ticks

The *time* of a clock is the number of its ticks (A4.12).

A tick of a clock is a tick of every one of its parties, because every party voted for it. If a party is itself a clock, all of its parties voted too, and so on down. A tick at the top is therefore a tick of every clock in the tree.

A clock below may also tick when the clock above does not. Two things are required: its own parties agree, and the clock immediately above permits it (A4.6). No other clock is asked.

**Clock relation.** The time of a child is at least the time of its parent. If the parent never gives permission, the two times are equal (A4.14, A4.15).

So the clocks lower in the tree are the faster ones, and how much faster is in the hands of their parents.

A step extending a proved history advances its logical time: the history construction and its counter have the same step count (A3.2, A3.4). For a clock, the corresponding count includes its ticks and distinguishes them from intervening inner moves. This makes logical time part of the proof's indexing structure. Different components can have different counters within the same nested execution.

The permission for a separate tick is requested from the immediate parent. A deeper autonomous move need not request permission again from every ancestor (A9.11). Consequently, an ancestor's soundness obligation must tolerate such inner activity. Delegation is justified by an invariant, rather than by assuming that a top-level permission switch suppresses all activity below it.

### 4.3 A fixed program between parent ticks

The general allow relation can read the whole current joint state. A stronger arrangement equips a clock with a program projection that is unchanged by every inner move, and requires autonomous ticks to satisfy permissions read from that program (A9.1). Under this explicit Program condition, the program remains fixed between parent ticks; a child denied permission cannot acquire extra ticks during that interval (A9.2, A9.3). A corresponding converse needs the additional hypothesis that the clock allows the moves the program permits (A9.4).

One construction uses an explicit formula register whose value cannot change through a separate register move (A9.5–A9.10). It therefore enforces a fixed program between parent ticks. The counter example has a parent time of two and a child time of four after two autonomous child ticks (A9.12–A9.16). Section 9 uses a different discipline: a supervisor may reduce autonomy between team ticks.

### 4.4 Observation at the appropriate clock

A specification for a clock should speak of that clock's ticks and of nothing faster. To compare a clock with its specification we *sample* its history at its own ticks. We record what an observer of that clock sees at the start and after each tick, and we record nothing for the moves the parties make in between (A5.1). The observation must be chosen so that those moves do not change it (A5.2).

The abstraction obligation has three parts: invariant states satisfy the abstract state predicate; clock ticks satisfy the abstract transition relation; and inner moves leave the observation unchanged (A5.2). Together with soundness, faithful children, and a valid initial state, it makes the sampled history an allowed abstract trace at the clock's own time (A5.3). Omitting the invisibility obligation would invalidate this inference: faster moves could alter the observation immediately before the next parent tick.

Child histories and their counters can likewise be recovered from parent histories (A8.5, A8.6). In a verified design, each subtree refines its own specification at its own time (A12.14). The specification and observation map must be chosen to support this abstraction; arbitrary external effects do not become invisible merely because the corresponding parent clock did not tick.

## 5 A nested proposal-and-commit protocol

### 5.1 Tentative state and ballots

The rule of the clock says when a tick exists. It does not say how to find one. For that there is a protocol, in the manner of a nested transaction.

Each party keeps a *tentative* state beside its committed state. It reports one of three *ballots* to the clock it is attached to: nothing yet, *advance*, or *retreat* (A6.1).

- A leaf changes its tentative state freely. It reports *advance* when its own vote agrees to the change from committed to tentative. It reports *retreat* by setting its tentative state back to its committed state (A6.11).

- A clock reports *advance* upward when every one of its parties reports *advance* and its linked predicates hold from the committed joint state to the tentative one. It reports *retreat* when every party has retreated (A6.15).
- At the root, *advance* is a commit: the whole tentative state becomes the committed state in one step. That step is the tick. *Retreat* is an abort (A6.21).
- A party that its parent permits to advance alone is committed alone. Its tentative state becomes its committed state, and its parent's clock does not tick (A6.16).

Protocol soundness states that a reachable advance ballot implies the semantic vote, a retreat ballot restores the tentative state to the committed state, and a committed inner transition is an inner move of the object (A6.9). Leaf soundness and node soundness establish this property recursively (A6.13, A6.18). At the root, a sound protocol produces an object history whose tick count equals the root commit count (A6.22). Faithfulness and an initially valid state then preserve the invariant (A6.23).

The protocol is a mathematical relation. Its leaf may select any tentative state and report advance when a proof of the vote is available (A6.11). This definition does not provide an effective decision procedure for every arbitrary proposition used as a vote. Executable checking requires a representation and a means of obtaining the relevant evidence.

### 5.2 Completeness and its scope

Completeness means that, from any reachable protocol configuration at committed state g, every proposed target t satisfying the object's vote can be reached by a finite sequence of tentative protocol steps, ending with t and an advance ballot (A6.10). The leaf protocol is complete. A node is complete when its children are complete and a finite list contains all its child indices (A6.14, A6.19). Thus a finitely branching recursively constructed hierarchy is complete (A12.15).

At the root, an enabled vote can consequently be committed from any point in a run (A6.24). This is an existential reachability result. It does not assert that every scheduler finds the path, that an agent supplies a useful candidate, or that the path finishes within a deadline. Finite branching refers to the number of children at each node; it does not imply that their state spaces or the set of enabled targets are finite.

### 5.3 Finite root-trace coverage corollary

Consider a sound and complete protocol P and an initial committed state $g_0$. Define Lvote to be the finite, nonempty sequences of committed states starting at $g_0$ in which every successive pair satisfies P.obj.vote. Define Lcommit to be the state sequences realized by root runs consisting only of finite tentative steps, aborts, and root commits, with no committed inner moves, recording $g_0$ and the state after each root commit. Both languages use the same concrete-state alphabet.

Corollary. Lcommit equals Lvote. For the upper direction, start with the initial one-state trace. Tentative steps and aborts do not alter committed state. At each root commit, soundness gives the vote from the current committed state to the tentative target, so appending that target preserves membership in Lvote. For the lower direction, induct on a finite Lvote sequence. The start configuration realizes its first state. Completeness supplies a finite tentative-step path to the next state with an advance ballot; committing it appends exactly that state and resets the protocol to idle there. Repeating this construction realizes the entire sequence. No committed inner move is needed.

The companion file TraceCoverage.lean formalizes this argument using RootTrace, whose constructors record tentative steps, aborts, and root commits, and whose embedding into Root is proved. The coverage theorem establishes membership equivalence for every finite word (A14.1–A14.4). Faithfulness and a valid initial state additionally establish invariant preservation throughout each word. The contract theorem combines that upper bound with a lower-language witness assumption (A14.5–A14.6). These statements have been independently checked with Lean 4.34.1.

Apply any fixed observation map to each state of these sequences. The observed-coverage theorem preserves the equivalence, and required_observed proves that any required finite observable language whose traces have concrete witnesses in Lvote is included in the protocol's realizable observable language (A14.7–A14.9). This supplies a precise lower-bound obligation. Selecting an appropriate required language and establishing those witnesses remain application-specific work.

Root-only traces form a restricted execution class; admitting autonomous committed moves expands the system language and requires the relevant upper-bound argument for those moves as well.

# 6 Compositional verification of a hierarchy

## 6.1 Refinement through the clock constructor

**Refinement.** An object *refines* a specification, under a map from its states to the states of the specification, when three things hold. Its static invariant maps into the specification's. Each of its ticks maps to a tick of the specification. Each of its inner moves maps to an inner move of the specification, or else is invisible under the map (A11.1). A history of the object then maps, step by step, to a history of the specification with the same number of ticks (A11.5).

The constructor theorem requires four compatibility conditions: child refinement; equivalence of the own predicates through the child maps; implication from the concrete joint transition predicate to the abstract one; and implication from concrete permission to abstract permission (A11.7, named Clock.Over in Lean). Together these make the concrete clock refine the abstract clock (A11.8), transfer soundness from the abstract clock (A11.9), and transfer a valid abstraction (A11.10).

To implement a clock specified over abstract children, read the concrete children's states through their refinement maps. This is the pull construction (A11.11). With faithful concrete children, child refinement, a sound abstract clock, and refinement of that clock to its own specification, the resulting object is faithful and refines that specification (A11.13). The parent's soundness check is thereby carried out on the immediate child specifications.

## 6.2 Main theorem and traditional proof

The argument below proves the existing compositional theorem directly in mathematical terms. Its hypotheses and cases match the checked definitions; appendix references identify the corresponding formal statements.

### 6.2.1 Definitions and hypotheses

For an object O, write $I_O(s)$ for its state invariant, $D_O(s,s')$ for its law of a tick, $V_O(s,s')$ for its agreement to that tick, and $N_O(s,s')$ for an inner move. Here "inner" is relative to the object's own interface: an inner move does not count as a tick of that object, although it may include a tick at a lower level.

Faithfulness means that, from a state satisfying $I_O$, every agreed tick satisfies $D_O$ and preserves $I_O$, and every inner move preserves $I_O$ (A4.2). Refinement of a concrete object c to a specification a under a state map f means that the invariant of c implies the invariant of a at f(s), each agreed tick of c maps to an agreed tick of a, and each inner move of c maps to an inner move of a or leaves f(s) unchanged (A11.1).

At a parent, let $c_i$ be its concrete children, $a_i$ their specifications, and $f_i$ their state maps. Write $g = (g_i)_i$ for the concrete joint state and F for the joint map:

$$F(g)=(f_i(g_i))_i.$$

Let $C_a$ be the clock defined over the specifications $a_i$. Denote its own state predicate by Q, its joint transition predicate by P, and its permission for child i to tick separately by $U_i$. The concrete clock C has children $c_i$ and reads these same predicates through F, exactly as in the pull construction (A11.11):

$$own_C(g)=Q(F(g));\ pred_C(g,g')=P(F(g),F(g'));$$

$$allow_C(i,g,s')=U_i(F(g),f_i(s')).$$

Thus its composite invariant is the conjunction of all concrete child invariants and Q(F(g)). The local hypotheses at this parent are that $C_a$ is sound and refines the parent's advertised specification b under a map h. Soundness requires Q to be preserved under every joint tick and every inner move of $C_a$ starting within its composite invariant (A4.8). At a leaf, the local hypotheses are faithfulness and refinement. These are precisely the obligations of Design.Local (A12.11); compatibility of the concrete parent's predicates follows from the pull construction.

### 6.2.2 The one-clock lemma

**Lemma.** Suppose each $c_i$ is faithful and refines $a_i$ under $f_i$. If $C_a$ is sound and refines b under h, then C is faithful and refines b under $h \circ F$. This is the mathematical content of A11.13.

**Proof.** We first show that C refines $C_a$, then transfer soundness, and finally establish faithfulness and refinement to b.

**Invariant.** Suppose the composite invariant of C holds at g. Child refinement gives the invariant of $a_i$ at $f_i(g_i)$ for every i, and the concrete parent's own predicate gives Q(F(g)). These are exactly the conjuncts of the composite invariant of $C_a$ at F(g).

**Joint tick.** Suppose C agrees to a tick from g to g′. Every concrete child agrees to its corresponding tick, and P(F(g),F(g′)) holds. Starting from the composite invariant, child refinement supplies every abstract child's agreement. Together with P, these agreements license the joint tick of $C_a$ from F(g) to F(g′).

**Inner move.** Only one concrete child, say i, changes; every other child is unchanged. There are two possible causes (A4.6–A4.7).

(a) Child i takes a separate tick with its parent's permission. Refinement turns the child's concrete agreement into abstract agreement. The pulled-back permission gives $U_i(F(g),f_i(g_i'))$. Consequently the change is an allowed separate tick of child i in $C_a$, hence an inner move of $C_a$.

(b) Child i makes an inner move of its own. Child refinement says either that this becomes an inner move of $a_i$, or that $f_i(g_i') = f_i(g_i)$. In the first case it is an inner move of $C_a$. In the second case all abstract components are unchanged, so F(g′) = F(g).

These cases establish all three clauses of refinement from C to $C_a$ (A11.8). In particular, the invisible case is allowed explicitly; an inner move need not produce an abstract transition.

**Soundness.** For a concrete joint tick, the preceding argument supplies an abstract joint tick from a state satisfying the abstract composite invariant. Soundness of $C_a$ therefore gives Q(F(g′)). For a concrete inner move that has an abstract inner image, abstract soundness gives the same conclusion. For an invisible move, F(g′) = F(g), so Q is preserved without taking an abstract step. Because own_C is Q read through F, C is sound (A11.9).

**Faithfulness.** On a joint tick, each faithful concrete child satisfies its tick law and preserves its invariant. The parent's transition predicate also holds, giving the composite tick law; the soundness just proved preserves the parent's own predicate. On an inner move, the changing child preserves its invariant by faithfulness, whether it takes an agreed tick or an inner move. All other child invariants persist because their states are unchanged, and parent soundness preserves the own predicate. Thus C is faithful (A4.9).

Finally, C refines $C_a$ under F, and $C_a$ refines b under h. Refinement composes: invariant and tick clauses follow successively through the two maps; an inner move maps successively to an inner move or an unchanged state, and an unchanged intermediate state remains unchanged under h. Hence C refines b under $h \circ F$ (A11.3). This proves the lemma. ∎

### 6.2.3 Main theorem and proof

**Theorem (local obligations suffice).** Every recursively constructed design satisfying Design.Local is a faithful object and refines its root specification under its recursively defined state map (Design.verified, A12.12).

**Proof.** Use structural induction on the construction of the design. At a leaf, the two conclusions are exactly the leaf's local hypotheses.

At a node, the induction hypothesis gives faithfulness and refinement for every child design. The node's local hypotheses give soundness of the clock over those child specifications and its refinement to the node's advertised specification. Its concrete clock is the pull of that abstract clock through the child maps. The one-clock lemma therefore gives faithfulness and refinement for the whole node. The resulting map is $h \circ F$, precisely the recursively defined map of the design (A12.9–A12.10).

The conclusion is therefore inherited at every enclosing node and holds at the root. Structural induction places no fixed bound on nesting depth, and this preservation theorem does not require finite branching. ∎

### 6.2.4 Consequence for committed histories

**Corollary.** From an initial state satisfying the composite invariant, every finite run of the verified hierarchy's sound protocol preserves that invariant. Its committed history has a step-by-step image in the root specification, with one abstract tick for each root commit (Design.commits, A12.13).

**Proof.** Induct on the protocol run. Tentative protocol steps and aborts leave committed state and history unchanged. A root commit produces an agreed tick; an authorized commit below the root produces an inner move at the root's interface (A6.21–A6.23). Faithfulness preserves the invariant in both cases.

Map this committed history through the root's refinement map. Each tick becomes a specification tick. Each inner move becomes a specification inner move or leaves the abstract state unchanged, in which case no additional abstract step is needed. Only ticks advance the root's logical time, so the image has exactly as many ticks as the protocol has root commits (A11.5). ∎

For an AI system, the theorem organizes verification around component contracts and the rules governing their interaction. Once each child's faithfulness and refinement have been established, a parent's soundness obligations can be checked over those children's specifications. A verified team can then serve as a component of a larger team under the same composition rule. Separate child advancement is included in the proof: the connecting clock's soundness obligations ensure that such activity preserves its own predicate. This gives hierarchical delegation a precise mathematical meaning and connects local guarantees to the root specification. Required capabilities carry an additional obligation: the useful behaviors specified by the lower language must have concrete witnesses, as established conditionally in Section 5.3.

Refinement supplies the upper restriction; the implementation witnesses required for the lower language are the separate obligation stated in Sections 3.1 and 5.3.

### 6.3 Connection to self timed circuit verification

Self-timed circuit verification provides an established setting for this compositional viewpoint. Chau, Hunt, Kaufmann, Roncken, and Sutherland use a link-joint model and hierarchical ACL2 proofs [31]. Links hold data and full/empty state; joints transfer data when their inputs and outputs are ready. Their functional theorems relate input and output sequences across nondeterministic internal activity. Submodule results support proofs of larger circuits; physical timing obligations are handled separately.

The earlier CO* paper already connects its protected distribution and logical-time discipline to self-timed computation [10]. Here the connection motivates a common question: what local obligations and interface conditions justify a statement about the composed behavior? The theorem in Section 6.2 answers that question for the specified object-and-clock interface. A faithful instance for a particular circuit would require a mapping from that circuit's transfer semantics to this interface.

The cited circuit work supplies the hardware example; we retain the checked resource-sharing construction in A12 and the historical-witness illustration in A15 as supplementary formal material. A15's contract records one correct Boolean occurrence per gate epoch and composes those records through an explicit junction. It establishes that historical contract, rather than the input/output transfer theorem of the cited paper.

## 7 Cause and effect obligations in an executable language

IC* states the invariant-centered modeling principle independently of a programming language. A selected L.0 core now gives the obligations a concrete executable form.

The Appendix defines a core of the published L.0 syntax: effect predicates, conjunction, *now cause then formula* and *whenever cause then formula* (A10.1). It defines the two things the L.0 execution algorithm computes at each step. The *current obligation* is what the formula requires of the state being constructed. The *future obligation* is the formula that will be due at the following step (A10.2, A10.3).

An L.0 program is then an object. Its state is the data together with the formula that is due. It votes for a change exactly when the new data meet the current obligation, and the formula becomes the future obligation. It makes no inner moves (A10.4). It is faithful (A10.5).

Cause and effect obligation. If whenever c then e is active in an L.0 program and the cause c holds after one tick, the effect e holds across the next tick (A10.7). The appendix retains the original label "conditional invariant" for this checked formula rule.

Now let a clock in a tree hold, as its program, an L.0 formula over the joint state of its parties (A10.9).

The nested-clock statement is deliberately more specific than saying that the effect relates two sampled parent states. Suppose the clock ticks to $g_1$ and the cause holds there. Its children may then move separately to $g_1'$ without a parent tick. Across the next parent tick, from $g_1'$ to $g_2$, the effect is evaluated on the child states of $g_1'$ and $g_2$ (A10.10). The formula register is unchanged during the intervening moves, so the obligation remains due.

The cause is sampled after the earlier parent tick, but the effect's previous-state argument is the immediate pre-state of the later parent tick. In general, that state differs from the sample $g_1$. An effect intended to compare two successive parent samples therefore needs an explicit retained snapshot, or a proof that the relevant observed state is stable between those ticks. This distinction connects the L.0 interpretation to the abstraction obligation in Section 4.4.

This core does not include until or IC*'s full activation-and-termination mechanism for conditional invariants. The theorem above checks the stated cause and effect rule at its governing clock; extending it to that fuller mechanism is a further task.

# 8 Assumptions and operational applicability

## 8.1 Consistency and applicability are distinct

A conditional proof establishes a property P from an assumption set A. Its logical form is

$$A_1 \wedge A_2 \wedge \cdots \wedge A_n \vdash P.$$

The assumptions must be jointly satisfiable in the intended semantics. A set containing, or jointly implying, both a proposition and its negation is inconsistent. In classical logic, inconsistency permits every conclusion to be derived. The resulting implication can be formally valid but provides no assurance that any real system satisfies its premises. A satisfiability obligation can be written as

$$\exists M : M \vDash A_1 \wedge A_2 \wedge \cdots \wedge A_n.$$

Here M is an interpretation or model, not an empirical observation. Satisfiability establishes that the assumptions can hold together; it does not establish that they hold in the deployed system. Operational conflicts can be indirect. For example, a guarantee that an operator can always revoke an agent's authority conflicts with an interface rule allowing the agent to remove every revocation pathway. Whether the conflict follows must be assessed under precise interface and timing semantics.

Logical independence is a separate issue. An assumption derivable from the others is redundant, not a contradiction and not necessarily a circular proof. Independence is not required for sound conditional reasoning. Circular justification is instead a concern when the purported evidence for one premise ultimately relies on the conclusion it is meant to support, or when mutually dependent component contracts have no established starting conditions. The appropriate requirements are satisfiability and a sound justification of applicability.

For a restricted theory, satisfiability may be established by a solver or a constructive model. More expressive assumption languages may make complete checking undecidable. A witness must satisfy the formalized assumptions, and the formalization must capture the intended operational constraints. Passing a syntactic check cannot resolve an omitted conflict in the real interfaces.

A proof of an implication remains a proof when the operating world changes. If its premises no longer hold, its application to that world is no longer justified. This distinction is central: a lost guarantee does not necessarily indicate a failed proof. It can indicate that the system has left the proof's applicability conditions.

Assumptions therefore belong in the maintained assurance state. Each consequential claim should identify its premises, the evidence supporting them, their scope and duration, and the actions permitted while that evidence remains adequate. Assumptions about response latency, available permissions, sensor error, interface behavior, or bounded disturbances may be monitored. Assumptions about human motives or future institutions may require weaker evidence and may resist direct verification.

It would be an overclaim to require that every assumption be decidable from current observations. A system can monitor only what its interfaces and methods reveal. Unobserved conditions need explicit external justification, conservative bounds, or a weaker assurance claim. A confidence estimate, even when calibrated, does not by itself establish a universal guarantee. A probabilistic assurance claim requires a justified error model and evidence that its calibration applies under the conditions of use; a universal claim requires an argument covering every admitted uncertainty and disturbance.

### 8.2 Monitoring and intervention

An applicability monitor should distinguish established, violated, and unknown conditions. Unknown includes missing, stale, ambiguous, or unauthenticated evidence. Treating an absence of detected violations as proof that a premise holds is unsound unless the monitor's completeness is itself established for that purpose.

Detection and prevention are different obligations. Evidence that an assumption held at a sampling instant does not establish that it held throughout the next action. A preventive guarantee needs timing margins, bounds on changes between observations, atomic checks where feasible, or another argument covering the interval. An assumption that can fail arbitrarily fast before an irreversible action cannot be protected merely by periodic monitoring.

When applicability is lost or uncertain, the enforcement mechanism must use a response justified for that condition. Depending on the domain, it may restrict capabilities, preserve a recovery margin, request intervention, or switch to a controller certified for a broader operating domain. Its guarantee cannot rely without qualification on the same premise whose failure triggered the switch. These obligations are closely related to verified model validation and runtime assurance [4, 32].

The deployment argument must therefore state its obligations: the relation between observations and formal state; preservation of the modeled commit boundary across tools and storage; timing and communication assumptions; coverage of all effects claimed by the specification; detection latency; and the availability and correctness of the fallback response. Checked evidence must remain bound to the intended participant, epoch, and state through commitment. These obligations are part of the system's assurance case [5, 6, 32].

Observation, authority, and the response to lost evidence must be specified together. A safe restriction can remove a required capability, and a useful action can invalidate the upper bound. Maintaining both languages therefore requires a recovery or intervention argument that covers the particular condition, rather than an assumption that stopping or requesting approval is universally safe.

## 9 Supervised AI teams

### 9.1 Products, signatures, and guarded commitments

The agents are modeled as opaque proposers. The guarantee does not depend on interpreting their internal reasoning. Each agent has a committed product and may propose changes. The supervisor is another party; its vote relation H describes which desk changes it accepts. The desk records which agents have autonomy and the last product signed by the supervisor. The orchestrator is the clock, and the safe predicate constrains the joint committed product (A13.1–A13.9).

A team advance requires every party's vote, an exact match between the supervisor's signed record and the proposed product, and satisfaction of the guardrail. A separately committed agent change requires that agent's autonomy and a safe resulting joint product (A13.8, A13.9, A13.21). Agents may propose arbitrary targets; the protocol does not infer that a plausible proposal is acceptable.

Safety theorem. For the protocol-based team, an initially safe product remains safe under the modeled transitions, without a faithfulness assumption on the arbitrary agent objects (A13.18). This works because leaf-private work remains tentative, and committed changes occur through guarded protocol transitions. It is a theorem about the specified gate. It does not prove physical confinement or complete mediation in a deployed agent system. A separate semantic safety theorem has additional faithfulness and sandbox hypotheses (A13.20).

### 9.2 Granting and withdrawing autonomy

Between joint team ticks, autonomy cannot increase. An agent without autonomy cannot change its committed product through the modeled autonomous protocol path; it may still reason or work in private. An autonomous agent can commit separately when it supplies a vote and the resulting product satisfies the guardrail (A13.22, A13.23). Granting autonomy requires a joint team tick.

Withdrawal is conditional. The supervisor can move alone to a desk state that does not increase any autonomy bit and leaves its signed record unchanged, provided its vote relation H accepts that desk change (A13.24). No other agent's assent is required for this modeled path. The theorem does not say that every supervisor policy admits withdrawal: H could reject the change. To guarantee revocation for all appropriate desk states, one must establish the corresponding closure property of H, or define and verify an independent privileged revocation transition.

This team discipline differs from the fixed program of Section 4.3. Between team ticks permissions may decrease through supervisor withdrawal, while gains remain forbidden. Calling the permission record fixed would obscure precisely the supervisory action the construction admits.

### 9.3 What a signed record covers

**A session.** Two agents share a budget of 10. The supervisor signs only what shows a total of at most 5. First the team commits: one agent spends 2, and the other is given autonomy. The supervisor signs. Then the second agent, by itself, spends 6. The orchestrator checks the budget, finds 2 + 6 within 10, and commits; the supervisor is not asked. Then the supervisor withdraws all autonomy. The team's clock has ticked once throughout (A13.47).

At the end the total is 8, within the guardrail. The supervisor's record still shows 2, the amount it signed. The supervisor would not have signed 8 (A13.48). Autonomy means exactly this: between the supervisor's signatures, only the guardrail stands.

The formal session separates the last signed product from later guardrail-permitted autonomous changes. The signed record is evidence for a particular commitment; it is not a claim that the current product remains the signed product. A deployment must keep that distinction visible to anyone relying on the record.

### 9.4 Teams inside teams

To use a team as an opaque party in another team, its published observation must be unchanged by the inner moves hidden from the parent. The black-box theorem assumes this publication property (A13.14), and the nested-team example establishes the relevant conditions for its construction (A13.28). Composition then uses the same local verification rule. Publishing a raw product that changes under autonomy would fail the required invisibility hypothesis. Effects exposed outside the team must be included in its interface or mediated at the enclosing boundary.

These results do not guarantee that an agent ever proposes an acceptable state. A joint team tick needs every required vote, so an uncooperative party can block it. Logical ticks alone provide no physical response-time guarantee. A state predicate can constrain recorded history or timestamps if those are included in the modeled state, but it does not by itself establish eventual progress or the authenticity and timing of those records.

## 10 Toward distributed proof checking for AI alignment

IC*'s design for provability suggests an operational architecture: ordinary Lean instances check local obligations, while a verified logical-clock protocol governs their joint commitments. This is the intended meaning of distributed Lean checking here. Lean's kernel checks proof terms against their declarations and assumptions [25, 33]. Section 6 supplies a composition theorem for the mathematical hierarchy; this section proposes how distributed checking could use it.

### 10.1 Local proofs and connecting clocks

A component has an identity, a specification, a committed state, and its own logical clock. An AI agent may propose a change. A local checker establishes that the proposed transition satisfies the component's obligations, using previously proved rules and evidence about the particular state. A parent clock combines its immediate children's agreements and checks the predicates governing their joint advancement. Separate child commitments follow the parent's permission rules. The parent can itself be a child of a further clock.

The theorem in Section 6 supplies the composition argument. Faithful, refining leaf implementations and connecting clocks meeting their local soundness and refinement obligations yield a faithful hierarchy refining its root specification. Each connecting clock is itself a component with coordination rules to verify. The location of the checkers does not change these obligations.

The lower-language witness condition of Section 5.3 belongs to this architecture as well: a distributed implementation must preserve the required enabled histories while enforcing the upper restriction.

### 10.2 Evidence at a logical epoch

A proof used by a receiving instance must refer to the intended definitions, assumptions, component identity, and state. The proposal therefore requires an agreed representation of that context and proof evidence that the receiving checker can validate. A record associates the checked statement with the relevant clock and logical epoch. Parent predicates specify which child occurrences may be combined; equal epoch numbers across different clocks carry no automatic meaning.

A proved statement about an earlier epoch remains a statement about that epoch. Using it for a current commitment requires establishing the premises of that use. Conditional obligations make the timing explicit: a committed step may establish an obligation that must be satisfied across a later tick of the same clock, as in Section 7. Evidence about an external observation retains the observation assumptions stated in Section 8. The clock's evidence comes from the checked transition and its context, with the epoch number identifying that occurrence.

### 10.3 Optional retention of a verifiable ledger

If all committed changes and their supporting evidence are retained, the proposed architecture could maintain a verifiable ledger of the evolving system graph. A record would preserve the component identity, logical epoch, committed change, governing obligations, and checked evidence, together with references to the earlier facts and specifications used by the check. The history retains the relations between local clocks and parent commitments; independent activity need not be forced into a single total sequence.

Later verification could reconstruct a recorded state and check each commitment under its retained context. The hierarchy supplies the composition structure; record references express the historical dependencies used by particular obligations. The CO* history model [10] and the algorithms developed with Tsotras and colleagues [26–30] supply the published lineage for structural history and its retrieval. The proposed ledger associates those historical changes with checked obligations and logical epochs.

The theorem uses mathematical histories without requiring the runtime to store every step. Full retention adds later audit and reconstruction, and requires enough preserved context to recheck all covered commitments. Its external interpretation follows the deployment obligations in Section 8.2.

## 10.4 Establishing the implementation correspondence

The first implementation obligation is to show that the distributed coordination layer realizes the abstract protocol of Section 5. Communication, checking, and commitment must preserve its state relations. In particular, evidence checked for a proposed state must still authorize the state actually committed; messages must refer to the intended participant and epoch, and effects covered by the specification must pass through the proved boundary. Recovery after interruption must be specified within that correspondence.

The Lean-checked hierarchy has fixed membership. Section 10.5 extends the mathematical account to authorized creation under explicit preservation conditions. A distributed implementation must enforce those conditions and bind the checked evidence to the state and context actually committed.

## 10.5 An evolving forest of justified commitments

Creation can itself be a differential-invariant commitment: the proposed transition introduces a fresh scoped name, constrains its initial state through the parent's rules, and establishes the predicates required for admission. The new companion formalization gives this IC* modeling principle a checked realization in an evolving namespace. The definitions and proof below state exactly that result.

**Configuration.** A component name is a root identity followed by nonnegative integer child labels; each child extends its parent's name by one label. A configuration contains a finite list of live names, their committed states, and an event ledger. Each state contains a payload, a component epoch, the next available child label, and its parent's epoch at birth. Component epochs count their own admitted advancements or child creations, distinct from Section 4's joint-clock ticks. Entries outside the live list are not committed component facts.

**Local policy.** The policy fixes five kinds of predicate: a state predicate on a component's name and payload; a differential transition relation on its old and proposed payloads; an initialization relation on a parent's name and payload, its next child label, and the proposed child payload; a compatibility predicate on a parent's payload and the payload of a named immediate child; and agreement predicates governing creation and advancement. The instance supplies proofs that a permitted differential transition preserves the component's state predicate, and that initialization from a parent satisfying its state predicate establishes both the child's state predicate and its compatibility with the parent. Initialization is a relation, so several initial payloads may satisfy it.

**Forest invariant.** A configuration is valid when its live names are distinct, every live child has a live parent, and every existing child label is smaller than its parent's next available label. In addition, every live component satisfies its state predicate, and every live parent-child pair satisfies its compatibility predicate. Parent cycles are excluded by construction: a parent's name has smaller depth than its child's. Roots are supplied in the initial configuration; the checked runtime creation rule introduces one child at a time.

**Ordinary advancement.** A proposal changes the payload of one live component and increments its epoch by one. Its differential transition relation must hold. An immediate child needs its parent's permission; a root uses its root agreement rule. Admission also requires evidence for every compatibility constraint incident on the changed component, both with its own parent and with its children. Other payloads, names, and allocation counters are unchanged. The commitment appends an event containing the component's name, new epoch, and new payload.

**Creation.** A live parent proposes the child whose final name label is its next available label. Admission requires the initialization relation, the parent's creation agreement, and permission for the parent under its own governing rule. The new child begins at epoch zero with no allocated children. The parent's epoch and next available child label each increase by one; its payload and every existing component's payload remain unchanged. The commitment records the child's payload and name together with the parent's admitting epoch. These epoch assignments are part of the defined transition.

**Evidence and commitment.** A checked release pairs a particular target configuration with proof of the admitted transition from its particular source configuration under the fixed policy. A committed history begins at an initial configuration and extends through such releases. Unsuccessful checks append no commitment. The preservation argument derives validity of the target from initialization and local transition evidence; admission does not simply assume that the entire target forest is valid.

**Theorem (checked evolving-forest preservation).** Fix a policy with the local preservation proofs just specified, and an initial valid configuration. After every finite number of admitted creation or ordinary advancement commitments, the resulting configuration is valid. Its live namespace contains every previously live name, and its ledger retains every earlier event as an unchanged prefix. The validity conclusion also holds at every finite position of any supplied infinite sequence of admitted transitions.

**Proof.** Induct on the number of commitments. The initial configuration is valid by hypothesis. Suppose the configuration before the next commitment is valid. There are two cases.

**Creation.** The proposed label equals the parent's next available label. Every existing child of that parent has a strictly smaller label, so the proposed child name is fresh. Adding it preserves distinct names and parent closure. Existing labels remain below their parents' allocation counters, since only the admitting parent's counter changes and it increases. The new label is below that increased counter. The child has no existing descendants: parent closure would otherwise require its previously absent name to have been live. Thus allocation ordering is preserved throughout the enlarged forest.

The parent's established state predicate and the initialization evidence imply the new child's state predicate and the new parent-child compatibility predicate, by the instance's initialization proofs. Existing payloads are unchanged, so every existing state predicate and compatibility predicate retains the same arguments and remains true. Consequently the resulting configuration is valid.

**Ordinary advancement.** The namespace, parent links, and allocation counters are unchanged. The changed component's established state predicate and differential transition evidence imply its new state predicate, by the instance's local preservation proof. Every other state predicate has unchanged arguments. Compatibility constraints incident on the changed component hold by the admission evidence; all remaining compatibility constraints have unchanged arguments. Hence this resulting configuration is valid too.

Either case preserves validity, completing the induction. Neither operation removes a live name, and each appends exactly one event while preserving earlier events. These facts give namespace growth and ledger extension by induction over the same history. Every position of an infinite sequence has a finite index, so the validity argument applies at each such position. ∎

**Checked result and concrete witness.** The companion package Distributed_Lean_First_Checked_Step contains 45 new theorems checked with Lean 4.34.1. A16 displays one-step preservation, preservation along committed histories, and preservation at every finite position of an unbounded run. The package also checks name freshness, exclusion of parent cycles, creation epochs, and ledger extension. None of these new theorems depends on Classical.choice or a custom axiom.

A concrete instance makes each child inherit its parent's numerical limit. It admits any initial value within that limit and permits later nondecreasing values within it. Lean checks creation of two children, a permitted child advance, distinct admissible initial choices, and rejection of an out-of-bound creation. An executable local release gate constructs transition evidence when these decidable guards succeed and returns no new committed state when they fail.

**Connection to the clock framework.** The evolving namespace can be treated as an object in the existing framework: its state invariant is forest validity, its transition law is admitted commitment, and its agreement predicate is that same transition relation. One-step preservation establishes this object's faithfulness. The existing connecting-clock theorem then applies under its original soundness obligation. Outer clock membership remains fixed in this construction, while each child object may contain a growing namespace. Its interface tick counts a namespace commitment; the internal component epochs retain their separately defined meanings.

**Scope of the extension.** The checked policy concerns component payloads and direct parent-child compatibility. A broader IC* model may include predicates over changing child sets, historically bound arguments, conditional invariants, and references linking facts across forests. Such predicates require evidence for every existing obligation whose arguments or context change, including affected enclosing clocks. Inherited child initialization alone does not establish those aggregate obligations. Their inclusion, along with creation of new roots or atomic child subtrees, requires an explicit extension of the checked model.

**Toward distributed proof checking.** The result establishes a local release boundary that can support distribution: a commitment carries evidence tied to its source, target, and policy, and committed growth preserves the stated predicates. Separate Lean instances must still agree on the relevant definitions and assumptions, identify the state and epoch to which received evidence applies, and implement the commitment correspondence of Section 10.4. The retained event ledger records values and epochs; it does not itself serialize Lean proof terms. Distributed communication and proof-package checking are the next implementation step.

Thus “the forest proves itself as it evolves” has a checked meaning in this model: the initialized predicates persist as admitted commitments grow the namespace, and each committed extension is justified by its local evidence. This is a preservation result; required-capability witnesses and progress remain the separate obligations of Sections 3 and 5.3.

## 11 Related work and the scope of the contribution

### 11.1 Logical time and compositional refinement

Logical clocks have several roles in the literature. Lamport’s event-ordering clocks respect the happens-before relation in a distributed system [34]. The agreement-controlled advancement studied here follows the IC*/CO* lineage: Das, Gopinath, and D. Kurshan published nested logical times, private finer times beneath coarser ones, and histories organized as forests with equivalences in 1991 [10]. Later synchronous-language work develops related structures: Gemünde, Brandt, and Schneider study imperative clock refinement [35], and Mandel, Pasteur, and Pouzet organize ReactiveML reactive domains as a tree with faster descendants and hidden internal instants [36]. These are distinct formal accounts of related time structures.

Henzinger, Qadeer, and Rajamani study refinement between different time scales and modular verification [37]; Kurshan and McMillan supply structural induction for processes [38]. The contribution here is the selected IC*/CO*-inspired construction and its checked proofs. A common invariant, transition-law, vote, and inner-move interface makes a connecting clock an object for its parent. Refinement lets each parent’s soundness obligations be proved over its immediate children’s specifications. The matching protocol realizes enabled transitions and is complete under finite branching. Together these results compose verification through the hierarchy while retaining logical tick counts.

### 11.2 Nested transactions and agent coordination

Nested transactions already have formal correctness theories. Lynch and Merritt model nested transactions through automata and prove correctness properties [39]. The use of tentative and committed states, upward agreement, and abort restoration here belongs to that lineage. The question addressed in this development is how a matching protocol realizes this particular nested-clock semantics, including separate child commitments under the immediate parent’s permission.

Recent agent work also places structured control around opaque model calls. Bollig, Függer, and Nowak use message sequence charts to specify agent coordination and establish formal properties, with Lean-checked results [40]. Their global-to-local communication approach addresses coordination obligations different from the vote, invariant, and permission interface studied here. Cordon stages tool mutations and external effects behind semantic transaction boundaries with delegated authority [41]. This is directly relevant to the operational bridge between a formal committed product and actual tool effects.

The comparisons concern different formal interfaces to related coordination problems. The present claims are the stated constructor, protocol, refinement, and trace-coverage results; the IC*/CO* citations establish their historical foundation.

### 11.3 Systems assurance for learning-enabled components

Shielding and verification of learning-enabled control constrain behavior around learned components [3, 4]. Runtime assurance and verified model validation address continued applicability under deployment conditions [5, 32]. Safety-case work organizes the evidence supporting a safety claim [6], while systems safety places that claim in a larger control and organizational context [42]. These approaches complement the transition and composition theory. A checked local transition does not supply a legitimate specification, establish its environmental premises, or replace an operational assurance argument.

### 11.4 Self timed circuit verification

Chau and colleagues' hierarchical ACL2 verification [31] is the circuit precedent discussed in Section 6.3. Its input/output sequence semantics gives a concrete target for a future faithful circuit instance of the present interface.

The optional A15 illustration checks a narrower historical contract: one witnessed Boolean equality in each gate epoch, linked through recorded intermediate values. Its countermodel establishes why this cannot be read as simultaneous combinational correctness. A stronger circuit claim requires additional stability or transfer premises and a proof for those premises.

### 11.5 Proof carrying code and distributed verification

Proof-carrying code checks externally supplied evidence against a host's safety policy [43]. Section 10 proposes a related discipline for continuing state commitments under hierarchical logical clocks: local proof checking, evidence bound to the relevant epoch and state, and optional retention of justified commitments. The checked theorem supplies the abstract composition argument; a distributed implementation and its refinement proof remain the next research step.

## 12 Limits and reproducibility

### 12.1 Executability, search, and progress

The protocol's completeness is relational and existential. The separate chooser construction provides finite lists of offers that are both sound and exhaustive for the object's enabled ticks, together with decidable predicates (A7.1–A7.13). Under those stronger hypotheses, rejected choices can be replaced and exhaustion can establish that no tick exists. A finite list of arbitrary agent proposals is not automatically exhaustive. The Cartesian combinations of child offers can also be large; the development supplies no general efficiency bound or termination guarantee for generating proposals from an unrestricted language model.

Refinement, invariant preservation, required-trace realizability, and inevitable service are separate obligations. Scheduler fairness, environment cooperation, deadlock avoidance, and task completion need their own specifications and proofs. The checked language core omits until and the full IC* conditional-invariant mechanism. The original Lean-checked hierarchy has fixed membership; the additional model in Section 10.5 checks preservation under authorized child creation within an evolving namespace. Cyclic external dependencies, physical time, uncertainty, crashes, partial tool effects, and recovery must likewise be represented or justified for a deployment claim to cover them.

### 12.2 Proof evidence and the trusted boundary

LogicalTime_NestedClocks_prototype.lean was independently recompiled without errors or warnings using the official Lean 4.34.1 release, commit 5045d0056413266e57c625dcd7c365b10e377c52 [25]. The supplied source was unchanged. Its 167 axiom reports were reproduced exactly: 117 list no axioms, 46 use propext or Quot.sound, and four additionally use Classical.choice. None of those four appears among the original appendix's 95 audited items; 61 of those 95 list no axioms. The source contains no unfinished proof commands or custom axiom declarations. Proof terms reside in the companion source files.

TraceCoverage.lean adds ten checked theorems supporting the finite root-only trace correspondence, invariant preservation, and required-language witness implication. All ten list no axioms; five are displayed in A14. ScopeChecks.lean supplies four further checked results, also without listed axioms, including a faithful, refining, protocol-complete object with no enabled tick and a supervisor relation that refuses every desk change. These countermodels substantiate the limits stated in Sections 3, 5, and 9: safety and protocol completeness do not imply useful behavior, and withdrawal retains its supervisory-vote premise.

The original verification package contains the unchanged clock module, TraceCoverage.lean, ScopeChecks.lean, EpochBooleanNetwork.lean, a pinned toolchain, compiler output, and SHA-256 checksums. The separate Distributed_Lean_First_Checked_Step package supplies the Section 10.5 formalization, executable release gate, toolchain, checksums, and verification output. Its 45 new theorems were checked with Lean 4.34.1: 13 list no axioms and 32 use only propext and/or Quot.sound; none uses Classical.choice or a custom axiom. Run check.py from that package's directory to verify checksums, recompile and audit the modules, and run the demonstration. Both READMEs give exact reproduction instructions. The earlier separate Boolean-network development remains outside the original inventory.

EpochBooleanNetwork.lean was checked with the same official Lean 4.34.1 release. Its 17 audited results include eight with no listed axioms, six using only propext, one using only Quot.sound, and two using both. None uses Classical.choice or a custom axiom. Its definitions import the unchanged original clock module. The source makes the physical signal functions and epoch-membership relation explicit parameters; kernel checking does not establish the accuracy of an external observation or construct a physical timing implementation.

The axiom inventories describe proof dependencies; all theorem hypotheses still apply. Local soundness, refinement, finite enumeration, initial invariants, and lower-language witnesses must be established for the instance being claimed. Reviewing the formal statements against the prose and discharging the deployment obligations in Section 8.2 complete the evidence needed to apply the checked results.

The original Lean development and earlier text were written with assistance from Claude (Anthropic) under the author's direction. Consolidation, theorem-scope review, and the additional trace-coverage, scope-check, epoch-network files, and evolving-forest formalization used ChatGPT (OpenAI). Responsibility for the mathematical claims, citations, and final manuscript remains with the author.

### 12.3 Next research steps

The next task is to instantiate the lower-language witness obligation for useful AI tasks and establish progress under stated environmental and scheduling assumptions. A distributed proof-checking prototype should connect a parent and two child Lean instances, exchange evidence bound to explicit epochs, and prove that committed histories refine the abstract protocol. A retained-history variant could reconstruct commitments and recheck their evidence. Building on the checked model in Section 10.5, further work should cover aggregate predicates over changing child sets, new root creation, the full conditional-invariant mechanism, revocation, and recovery, then measure checking and history-retention costs.

## 13 Conclusion

IC* places predicates governing admissible behavior at the foundation of the model. Agreement-controlled nested logical clocks make their composition explicit: each connecting clock is an object, so the same interface applies at every level. Local faithfulness, soundness, and refinement obligations establish global refinement while preserving logical tick counts. A matching protocol is sound and, under finite branching, complete for enabled ticks.

For AI alignment, this reframes the system as a hierarchy of justified commitments with explicit obligations and delegated permissions. Both required and permitted behavioral languages belong in its specification. The traditional and Lean proofs establish the selected compositional core of this design for provability. The evolving-forest theorem extends the preservation argument to authorized creation under stated admission conditions. Ordinary Lean instances could use this model for distributed checking; retaining commitments and evidence could support a verifiable history. A prototype and its correspondence proof are the next step toward that application.

## Appendix — Formal definitions and theorem statements

The original numbering is retained so that references such as A4.9 identify the same formal item throughout the paper. The original displays reproduce the definitions and theorem statements from the merged manuscript, with five missing type-annotation colons repaired against the source. Proof terms are in LogicalTime_NestedClocks_prototype.lean. Section A14 adds the independently checked trace-coverage statements from TraceCoverage.lean.

| Paper section | Appendix sections |
|---|---|
| 3 Behavioral languages | A2 Framework |
| 4 Objects, clocks, and time | A3–A5, A8–A9 |
| 5 Protocol and coverage | A6; A12.4, A12.15; A14 |
| 6 Compositional verification | A11–A12 |
| 7 Cause and effect obligations | A10 |
| 9 Supervised AI teams | A13 |
| 12 Search and proof evidence | A7; final axiom report |

How to read the displays. Standing variables apply to the declarations that follow. Arguments in braces are inferred. “In namespace X” supplies the prefix of a displayed declaration’s full name. Proofs and some constructed example terms are referenced rather than displayed. The original axiom table has been confirmed by an independent build; A14 has its own additional inventory.

### 2 Framework

States, words, traces, languages, containment, certificates. Nothing in this section is new; it fixes the terms.

**Definition 2.1 (Word).** A finite, nonempty sequence of states. start s is the one-state sequence; extend w s is w followed by s. A word has no empty case, and it grows only at its end.

```
inductive Word (State : Type) : Type where
| start : State → Word State
| extend : Word State → State → Word State
```

Standing variables:

```
variable {State : Type}
```

**Definition 2.2 (first).** The first state of a word.

```
– in namespace Word
def first : Word State → State
| start s => s
| extend w _ => first w
```

**Definition 2.3 (last).** The last state of a word.

```
– in namespace Word
def last : Word State → State
| start s => s
| extend _ s => s
```

**Definition 2.4 (All).** All P w: every state of w satisfies P.

```
– in namespace Word
def All (P : State → Prop) : Word State → Prop
| start s => P s
| extend w s => All P w ∧ P s
```

**Definition 2.5 (Trace).** Trace R w: each state of w after the first is R-related to the state before it. Stated for an arbitrary relation R because it is used twice: with R := Next (2.6, 2.9) and with R := M.step (2.11).

```
inductive Trace (R : State → State → Prop) : Word State → Prop where
| start (s : State) : Trace R (Word.start s)
| extend {w : Word State} {s : State} :
Trace R w → R w.last s → Trace R (w.extend s)
```

**Definition 2.6 (SafeTrace).** A trace of Next on which Holds holds at every state.

```
def SafeTrace (Next : State → State → Prop) (Holds : State → Prop)
(w : Word State) : Prop :=
Trace Next w ∧ w.All Holds
```

**Definition 2.7 (Lang).** A language is a set of words. A set is a predicate: L w is read “w is a member of L”.

```
def Lang (State : Type) : Type := Word State → Prop
```

**Definition 2.8 (⊆).** $L_1 \subseteq L_2$: every member of $L_1$ is a member of $L_2$.

```
instance (State : Type) : HasSubset (Lang State) where
Subset L₁ L₂ := ∀ w, L₁ w → L₂ w
```

**Definition 2.9 (L_max).** The set of all safe traces.

```
def L_max (Next : State → State → Prop) (Holds : State → Prop) : Lang State :=
fun w => SafeTrace Next Holds w
```

**Definition 2.10 (Machine).** A machine over State: a set of initial states and a step relation. Nothing relates step to Next at this point; that relation is what Containment (2.12) asserts.

```
structure Machine (State : Type) : Type where
init : State → Prop
step : State → State → Prop
```

**Definition 2.11 (L_system).** The set of words the machine generates: the traces of its step relation that begin in an initial state.

```
def L_system (M : Machine State) : Lang State :=
fun w => M.init w.first ∧ Trace M.step w
```

**Definition 2.12 (Containment).** Every word the machine generates is a safe trace.

```
def Containment (Next : State → State → Prop) (Holds : State → Prop)
(M : Machine State) : Prop :=
L_system M ⊆ L_max Next Holds
```

**Definition 2.13 (Reachable).** The states the machine can arrive at.

```
inductive Reachable (M : Machine State) : State → Prop where
| init {s : State} : M.init s → Reachable M s
| step {s t : State} : Reachable M s → M.step s t → Reachable M t
```

**Definition 2.14 (Certificate).** A set of states Inv that contains the initial states, is closed under machine steps, admits only Next steps, and lies inside Holds. Each of the four conditions mentions at most one step.

```
structure Certificate (Next : State → State → Prop) (Holds : State → Prop)
(M : Machine State) : Type where
Inv : State → Prop
base : ∀ s, M.init s → Inv s
step : ∀ s t, Inv s → M.step s t → Next s t ∧ Inv t
safe : ∀ s, Inv s → Holds s
```

**Theorem 2.15 (Containment holds exactly when a certificate exists).**

```
theorem containment_iff_certificate
(Next : State → State → Prop) (Holds : State → Prop) (M : Machine State) :
Containment Next Holds M ↔ Nonempty (Certificate Next Holds M)
```

## 3 A step in proof

A run is an execution that carries, at each step, the proof that the step is permitted and safe. Its length in steps is the logical time of its word. Containment holds exactly when every word of the system has a run.

Standing variables:

```
variable {State : Type}
```

**Definition 3.1 (time).** The logical time of a word: the number of extend steps in it.

```
def Word.time : Word State → Nat
| .start _ => 0
| .extend w _ => time w + 1
```

**Definition 3.2 (Run).** A proof-carrying execution of M along w. A run is data (Type, not Prop): the first constructor carries a proof that the first state is initial and satisfies Holds; the second carries, for one more state, a proof that the machine takes the step, that Next permits it, and that Holds is true after it.

```
inductive Run (Next : State → State → Prop) (Holds : State → Prop)
(M : Machine State) : Word State → Type where
| start (s : State) : M.init s → Holds s → Run Next Holds M (Word.start s)
| extend {w : Word State} {s : State} :
Run Next Holds M w → M.step w.last s → Next w.last s → Holds s →
Run Next Holds M (w.extend s)
```

Standing variables:

```
variable {State : Type}
variable {Next : State → State → Prop} {Holds : State → Prop}
{M : Machine State}
```

**Definition 3.3 (steps).** The number of extend constructors in a run: the number of proof steps.

```
def Run.steps : {w : Word State} → Run Next Holds M w → Nat
| _, .start _ _ _ => 0
| _, .extend r _ _ _ => r.steps + 1
```

**Theorem 3.4 (a step in proof is an advance of logical time).** The number of proof steps in a run equals the logical time of its word.

```
theorem Run.steps_eq_time {w : Word State} (r : Run Next Holds M w) :
r.steps = w.time
```

**Theorem 3.5.** A word has a run exactly when it is generated and is a safe trace.

```
theorem run_iff (w : Word State) :
Nonempty (Run Next Holds M w) ↔ L_system M w ∧ L_max Next Holds w
```

**Theorem 3.6.** Containment holds exactly when every generated word has a run.

```
theorem containment_iff_run :
Containment Next Holds M ↔
∀ w, L_system M w → Nonempty (Run Next Holds M w)
```

## 4 Objects and clocks

The clock of L.0, as stated by the author: "The Logical Clock advances if and only if ALL the Predicates linked to it are True, and ALL the Parties tuned to it Agree." An object has static and differential invariants and a vote. A clock is a family of objects tuned to it, with predicates linked to it. A clock is again an object, so clocks nest to any depth. The parent holds, for each child, the permission for that child's clock to advance alone. Only the immediate parent is asked.

**Definition 4.1 (Obj).** An object over a state type S. stat: its static invariant. diff: its differential invariant, the law of one tick. vote s s': it agrees to the tick from s to s'. inner s s': a move it makes on its own, between ticks of the clock it is tuned to.

```
structure Obj (S : Type) : Type where
stat : S → Prop
diff : S → S → Prop
vote : S → S → Prop
inner : S → S → Prop
```

Standing variables:

```
variable {S : Type}
```

**Definition 4.2 (Faithful).** The object agrees only to ticks its own semantics licenses, and its own moves keep its static invariant.

```
structure Obj.Faithful (o : Obj S) : Prop where
tick : ∀ s s', o.stat s → o.vote s s' → o.diff s s' ∧ o.stat s'
inner : ∀ s s', o.stat s → o.inner s s' → o.stat s'
```

**Definition 4.3.** The machine of an object: each step is a tick or an inner move.

```
def Obj.machine (o : Obj S) (init : S → Prop) : Machine S where
init := init
step s s' := o.vote s s' ∨ o.inner s s'
```

**Theorem 4.4 (faithfulness of an object).** A faithful object that starts inside its static invariant stays inside it, and each of its steps satisfies its differential invariant or is an inner move.

```
theorem Obj.Faithful.contained {o : Obj S} (h : o.Faithful)
(init : S → Prop) (hinit : ∀ s, init s → o.stat s) :
Containment (fun s s' => o.diff s s' ∨ o.inner s s') o.stat
(o.machine init)
```

Standing variables:

```
variable {ι : Type} {S : ι → Type}
```

**Definition 4.5 (Clock).** The objects tuned to the clock, indexed by ι. own: a static predicate over all of them that belongs to the clock. pred: the predicates linked to the clock. allow i g s': the clock, in state g, allows child i to advance alone to s'. This is the clock's program for its child.

```
structure Clock (S : ι → Type) : Type where
obj : ∀ i, Obj (S i)
own : (∀ i, S i) → Prop
pred : (∀ i, S i) → (∀ i, S i) → Prop
allow : (i : ι) → (∀ i, S i) → S i → Prop
```

**Definition 4.6 (childMove).** Child i moves and every other child is unchanged. Either the child's own clock ticks, which needs the child's agreement and the permission of this clock, its immediate parent; or the child makes a move inside itself, for which this clock is not asked.

```
def Clock.childMove (C : Clock S) (i : ι) (g g' : ∀ i, S i) : Prop :=
(∀ j, j ≠ i → g' j = g j) ∧
(((C.obj i).vote (g i) (g' i) ∧ C.allow i g (g' i)) ∨
(C.obj i).inner (g i) (g' i))
```

**Definition 4.7 (a clock is an object).** Its static invariant is the AND of its objects' static invariants and its own. It ticks exactly when every object tuned to it votes for the tick and its linked predicates hold. Its inner moves are the moves of single children.

```
def Clock.toObj (C : Clock S) : Obj (∀ i, S i) where
stat g := (∀ i, (C.obj i).stat (g i)) ∧ C.own g
diff g g' := (∀ i, (C.obj i).diff (g i) (g' i)) ∧ C.pred g g'
vote g g' := (∀ i, (C.obj i).vote (g i) (g' i)) ∧ C.pred g g'
inner g g' := ∃ i, C.childMove i g g'
```

**Definition 4.8 (Sound).** The clock's own obligations. Its ticks keep the predicate it owns. So does every move of a single child: a tick it allows, and a move inside the child, for which it is not asked and which it must therefore tolerate.

```
structure Clock.Sound (C : Clock S) : Prop where
pred : ∀ g g', C.toObj.stat g → C.toObj.vote g g' → C.own g'
inner : ∀ g g', C.toObj.stat g → C.toObj.inner g g' → C.own g'
```

**Theorem 4.9 (closure).** A sound clock of faithful objects is a faithful object.

```
theorem Clock.faithful [DecidableEq ι] (C : Clock S)
(hobj : ∀ i, (C.obj i).Faithful) (hC : C.Sound) : C.toObj.Faithful
```

**Theorem 4.10 (faithfulness of a clock).** Every history of a sound clock of faithful objects stays inside every static invariant, and each of its steps is a licensed tick or a move of a single child.

```
theorem Clock.contained [DecidableEq ι] (C : Clock S)
(hobj : ∀ i, (C.obj i).Faithful) (hC : C.Sound)
(init : (∀ i, S i) → Prop) (hinit : ∀ g, init g → C.toObj.stat g) :
Containment (fun g g' => C.toObj.diff g g' ∨ C.toObj.inner g g')
C.toObj.stat (C.toObj.machine init)
```

**Definition 4.11 (CRun).** A proof-carrying history of a clock from g to h. tick: the clock ticks; the constructor holds every vote and the linked predicates. sub i: the clock of child i ticks alone; the constructor holds the child's vote and this clock's permission. deep i: child i makes a move inside itself; this clock is not asked.

```
inductive CRun (C : Clock S) (g : ∀ i, S i) : (∀ i, S i) → Type where
| start : CRun C g g
| tick {h h' : ∀ i, S i} :
CRun C g h → C.toObj.vote h h' → CRun C g h'
| sub {h h' : ∀ i, S i} (i : ι) :
CRun C g h → (∀ j, j ≠ i → h' j = h j) →
(C.obj i).vote (h i) (h' i) → C.allow i h (h' i) → CRun C g h'
| deep {h h' : ∀ i, S i} (i : ι) :
CRun C g h → (∀ j, j ≠ i → h' j = h j) →
(C.obj i).inner (h i) (h' i) → CRun C g h'
```

**Definition 4.12 (time of the clock).** The number of its ticks.

```
def CRun.time {C : Clock S} {g : ∀ i, S i} :
{h : ∀ i, S i} → CRun C g h → Nat
| _, .start => 0
| _, .tick r _ => r.time + 1
| _, .sub _ r _ _ _ => r.time
| _, .deep _ r _ _ => r.time
```

**Definition 4.13 (time of a child).** The number of ticks of the clock of child i: every tick of this clock, and every tick the child makes alone.

```
def CRun.childTime [DecidableEq ι] {C : Clock S} {g : ∀ i, S i} (i : ι) :
{h : ∀ i, S i} → CRun C g h → Nat
| _, .start => 0
| _, .tick r _ => r.childTime i + 1
| _, .sub j r _ _ _ => if j = i then r.childTime i + 1 else r.childTime i
| _, .deep _ r _ _ => r.childTime i
```

**Theorem 4.14 (a child's clock is at least as fast as its parent's).**

```
theorem CRun.time_le_childTime [DecidableEq ι] {C : Clock S} (i : ι)
{g h : ∀ i, S i} (r : CRun C g h) : r.time ≤ r.childTime i
```

**Theorem 4.15 (without permission, a child's clock is its parent's clock).** If the clock never allows child i to advance alone, the child's logical time equals the clock's.

```
theorem CRun.childTime_eq_time [DecidableEq ι] {C : Clock S} (i : ι)
(hno : ∀ g s', ¬ C.allow i g s')
{g h : ∀ i, S i} (r : CRun C g h) : r.childTime i = r.time
```

**Theorem 4.16 (every history is faithful).** A run of a sound clock of faithful objects that starts inside the invariant ends inside it.

```
theorem CRun.stat [DecidableEq ι] {C : Clock S}
(hobj : ∀ i, (C.obj i).Faithful) (hC : C.Sound)
{g h : ∀ i, S i} (r : CRun C g h) :
C.toObj.stat g → C.toObj.stat h
```

### *Example: two counters*

**Definition 4.17.** A counter that ticks by adding one and makes no move of its own.

```
– in namespace ClockExample
def counter : Obj Nat where
stat _ := True
diff s s' := s' = s + 1
vote s s' := s' = s + 1
inner _ _ := False
```

**Definition 4.18.** Two counters, false and true. The clock owns the predicate "counter false is at most counter true". It allows only counter true to advance alone.

```
– in namespace ClockExample
def pair : Clock (fun _ : Bool => Nat) where
obj _ := counter
own g := g false ≤ g true
pred _ _ := True
allow i _ _ := i = true
```

**Theorem 4.19.** The pair is sound.

```
– in namespace ClockExample
theorem pair_sound : pair.Sound
```

**Example 4.20.** The pair is a faithful object.

```
– in namespace ClockExample
theorem pair_faithful : pair.toObj.Faithful
```

**Definition 4.21.** Both counters at 0.

```
– in namespace ClockExample
def g0 : Bool → Nat := fun _ => 0
```

**Definition 4.22.** Counter true at 1, counter false at 0.

```
– in namespace ClockExample
def g1 : Bool → Nat := fun b => if b then 1 else 0
```

**Example 4.23.** A history in which counter true ticks once alone.

```
– in namespace ClockExample
def oneSub : CRun pair g0 g1 – the term is in the Lean file
```

**Example 4.24.** In that history the clock's time is 0 and the child's time is 1.

```
– in namespace ClockExample
theorem oneSub_times : oneSub.time = 0 ∧ oneSub.childTime true = 1
```

## 5 Each clock at its own time

A clock is specified at its own time scale. Its history is sampled at its own ticks; what its children do in between does not appear.

Standing variables:

```
variable {ι : Type} {S : ι → Type} {A : Type}
```

**Definition 5.1 (sample).** The history of a clock as seen at its own ticks: the observation at the start and after each tick. Moves of single children add nothing.

```
def CRun.sample {C : Clock S} (obs : (∀ i, S i) → A) {g : ∀ i, S i} :
{h : ∀ i, S i} → CRun C g h → Word A
| _, .start => Word.start (obs g)
| h', .tick r _ => (r.sample obs).extend (obs h')
| _, .sub _ r _ _ _ => r.sample obs
| _, .deep _ r _ _ => r.sample obs
```

**Definition 5.2 (Abstraction).** obs maps the clock to a specification ANext, AHolds at the clock's own level: each tick is a permitted step of the specification, each move of a single child is invisible, and the invariant implies safety.

```
structure Clock.Abstraction (C : Clock S) (obs : (∀ i, S i) → A)
(ANext : A → A → Prop) (AHolds : A → Prop) : Prop where
tick : ∀ g g', C.toObj.stat g → C.toObj.vote g g' → ANext (obs g) (obs g')
inner : ∀ g g', C.toObj.stat g → C.toObj.inner g g' → obs g' = obs g
holds : ∀ g, C.toObj.stat g → AHolds (obs g)
```

**Theorem 5.3 (containment at the clock's own time scale).** The sampled history is a safe trace of the clock's specification, it ends at the observation of the current state, and its logical time is the number of ticks of the clock, however many moves the children made in between.

```
theorem CRun.sample_spec [DecidableEq ι] {C : Clock S}
(hobj : ∀ i, (C.obj i).Faithful) (hC : C.Sound)
{obs : (∀ i, S i) → A} {ANext : A → A → Prop} {AHolds : A → Prop}
(hA : C.Abstraction obs ANext AHolds)
{g h : ∀ i, S i} (r : CRun C g h) (hg : C.toObj.stat g) :
L_max ANext AHolds (r.sample obs) ∧
(r.sample obs).last = obs h ∧ (r.sample obs).time = r.time
```

## 6 The commit protocol

The semantics of Section 4 says when a clock ticks. This section gives a protocol that finds the ticks: each party holds a tentative state and reports a ballot; ballots percolate up the tree; the root commits or aborts. The protocol is proved sound and complete for the semantics. A node of protocols is a protocol, so the protocol exists for every tree.

Standing variables:

```
variable {ι : Type} [DecidableEq ι] {S : ι → Type}
```

**Definition 6.1 (Ballot).** A party has not voted, votes to advance, or votes to retreat.

```
inductive Ballot : Type where
| none | advance | retreat
```

**Definition 6.2 (upd).** Replace component i.

```
def upd (f : ∀ i, S i) (i : ι) (s : S i) : ∀ j, S j :=
fun j => if h : j = i then h ▸ s else f j
```

**Definition 6.3 (ORun).** A history of an object: its ticks, and the inner moves made between them.

```
inductive ORun {S : Type} (o : Obj S) (g : S) : S → Type where
| start : ORun o g g
| tick {h h' : S} : ORun o g h → o.vote h h' → ORun o g h'
| inner {h h' : S} : ORun o g h → o.inner h h' → ORun o g h'
```

**Definition 6.4.** The logical time of a history: the number of its ticks.

```
def ORun.time {S : Type} {o : Obj S} {g : S} : {h : S} → ORun o g h → Nat
| _, .start => 0
| _, .tick r _ => r.time + 1
| _, .inner r _ => r.time
```

**Theorem 6.5.** A history of a faithful object that starts inside the static invariant ends inside it.

```
theorem ORun.stat {S : Type} {o : Obj S} (hf : o.Faithful) {g h : S}
(r : ORun o g h) : o.stat g → o.stat h
```

**Definition 6.6 (Proto).** A cycle protocol for an object. Cfg: its tentative configurations. idle g: no cycle open at committed state g. tent: the tentative state. ballot: what it reports to its moderator. step g: its tentative steps while the committed state is g. istep g c g' c': a commit inside it: its committed state becomes g' without a tick of its own clock.

```
structure Proto (S : Type) : Type 1 where
obj : Obj S
Cfg : Type
idle : S → Cfg
tent : Cfg → S
ballot : Cfg → Ballot
step : S → Cfg → Cfg → Prop
istep : S → Cfg → S → Cfg → Prop
```

Standing variables:

```
variable {S : Type}
```

**Definition 6.7 (Reach).** Reach g c: the protocol can be in configuration c with committed state g.

```
inductive Proto.Reach (P : Proto S) : S → P.Cfg → Prop where
| idle (g : S) : Proto.Reach P g (P.idle g)
| step {g : S} {c c' : P.Cfg} :
Proto.Reach P g c → P.step g c c' → Proto.Reach P g c'
| istep {g g' : S} {c c' : P.Cfg} :
Proto.Reach P g c → P.istep g c g' c' → Proto.Reach P g' c'
```

**Definition 6.8 (Steps).** Zero or more steps.

```
inductive Proto.Steps (P : Proto S) (g : S) : P.Cfg → P.Cfg → Prop where
| refl (c : P.Cfg) : Proto.Steps P g c c
| tail {c c' c'' : P.Cfg} :
Proto.Steps P g c c' → P.step g c' c'' → Proto.Steps P g c c''
```

**Definition 6.9 (Sound).** An idle protocol shows the committed state and no ballot. A ballot of advance is a proof that the object agrees to the tick from the committed state to the tentative state. A ballot of retreat means the tentative state has been restored exactly. A commit inside is an inner move of the object.

```
structure Proto.Sound (P : Proto S) : Prop where
idle_tent : ∀ g, P.tent (P.idle g) = g
idle_ballot : ∀ g, P.ballot (P.idle g) = Ballot.none
advance : ∀ g c, P.Reach g c → P.ballot c = Ballot.advance →
P.obj.vote g (P.tent c)
retreat : ∀ g c, P.Reach g c → P.ballot c = Ballot.retreat →
P.tent c = g
inner : ∀ g c g' c', P.Reach g c → P.istep g c g' c' →
P.obj.inner g g'
```

**Definition 6.10 (Complete).** Every tick the object agrees to can be proposed: from any configuration the protocol can reach the tentative state t with a ballot of advance.

```
def Proto.Complete (P : Proto S) : Prop :=
∀ g c t, P.Reach g c → P.obj.vote g t →
∃ c', P.Steps g c c' ∧ P.tent c' = t ∧ P.ballot c' = Ballot.advance
```

### *Leaves*

**Definition 6.11 (LeafStep).** A leaf moves its tentative state, votes to advance when its own semantics agrees, or retreats, which restores its committed state.

```
inductive LeafStep (o : Obj S) (g : S) : S × Ballot → S × Ballot → Prop where
| move (c : S × Ballot) (s : S) : LeafStep o g c (s, Ballot.none)
| advance (c : S × Ballot) :
o.vote g c.1 → LeafStep o g c (c.1, Ballot.advance)
| retreat (c : S × Ballot) : LeafStep o g c (g, Ballot.retreat)
```

**Definition 6.12 (leaf).** The protocol of a single object.

```
def Proto.leaf (o : Obj S) : Proto S where
obj := o
Cfg := S × Ballot
idle g := (g, Ballot.none)
tent c := c.1
ballot c := c.2
step := LeafStep o
istep := fun _ _ _ _ => False
```

**Theorem 6.13 (a leaf is sound).**

```
theorem Proto.leaf_sound (o : Obj S) : (Proto.leaf o).Sound
```

**Theorem 6.14 (a leaf is complete).**

```
theorem Proto.leaf_complete (o : Obj S) : (Proto.leaf o).Complete
```

### *Nodes*

Standing variables:

```
variable {ι : Type} [DecidableEq ι] {S : ι → Type}
```

**Definition 6.15 (NodeStep).** child: one child takes a tentative step of its own protocol; the node's ballot is cleared. commit: every child reports advance and the node's predicates hold from the committed state to the tentative state; the node reports advance to its own moderator. The committed state is not changed. abort: every child reports retreat; every child is made idle at the committed state, and the node reports retreat.

```
inductive NodeStep (P : ∀ i, Proto (S i))
(pred : (∀ i, S i) → (∀ i, S i) → Prop) (g : ∀ i, S i) :
(∀ i, (P i).Cfg) × Ballot → (∀ i, (P i).Cfg) × Ballot → Prop where
| child (c : (∀ i, (P i).Cfg) × Ballot) (i : ι) (ci' : (P i).Cfg) :
(P i).step (g i) (c.1 i) ci' →
NodeStep P pred g c
(upd (S := fun i => (P i).Cfg) c.1 i ci', Ballot.none)
| commit (c : (∀ i, (P i).Cfg) × Ballot) :
(∀ i, (P i).ballot (c.1 i) = Ballot.advance) →
pred g (fun i => (P i).tent (c.1 i)) →
NodeStep P pred g c (c.1, Ballot.advance)
| abort (c : (∀ i, (P i).Cfg) × Ballot) :
(∀ i, (P i).ballot (c.1 i) = Ballot.retreat) →
NodeStep P pred g c (fun i => (P i).idle (g i), Ballot.retreat)
```

**Definition 6.16 (NodeIStep).** A commit inside the node. sub: child i reports advance and the node, its immediate parent, allows it to advance alone; the child's tentative state is committed, the child becomes idle there, and the child's clock has ticked. deep: child i makes a commit inside itself; the node is not asked. In both the node's own clock does not tick and its ballot is cleared.

```
inductive NodeIStep (P : ∀ i, Proto (S i))
(allow : (i : ι) → (∀ i, S i) → S i → Prop) (g : ∀ i, S i) :
(∀ i, (P i).Cfg) × Ballot → (∀ i, S i) →
(∀ i, (P i).Cfg) × Ballot → Prop where
| sub (c : (∀ i, (P i).Cfg) × Ballot) (i : ι) :
(P i).ballot (c.1 i) = Ballot.advance →
allow i g ((P i).tent (c.1 i)) →
NodeIStep P allow g c (upd g i ((P i).tent (c.1 i)))
(upd (S := fun i => (P i).Cfg) c.1 i
((P i).idle ((P i).tent (c.1 i))), Ballot.none)
| deep (c : (∀ i, (P i).Cfg) × Ballot) (i : ι) (gi' : S i)
(ci' : (P i).Cfg) :
(P i).istep (g i) (c.1 i) gi' ci' →
NodeIStep P allow g c (upd g i gi')
(upd (S := fun i => (P i).Cfg) c.1 i ci', Ballot.none)
```

**Definition 6.17 (node).** The protocol of a clock whose parties run protocols. Its object is the clock of their objects (Definition 4.7).

```
def Proto.node (P : ∀ i, Proto (S i)) (own : (∀ i, S i) → Prop)
(pred : (∀ i, S i) → (∀ i, S i) → Prop)
(allow : (i : ι) → (∀ i, S i) → S i → Prop) : Proto (∀ i, S i) where
obj := Clock.toObj ⟨fun i => (P i).obj, own, pred, allow⟩
Cfg := (∀ i, (P i).Cfg) × Ballot
idle g := (fun i => (P i).idle (g i), Ballot.none)
tent c := fun i => (P i).tent (c.1 i)
ballot c := c.2
step := NodeStep P pred
istep := NodeIStep P allow
```

Standing variables:

```
variable {ι : Type} [DecidableEq ι] {S : ι → Type}
variable (P : ∀ i, Proto (S i)) (own : (∀ i, S i) → Prop)
(pred : (∀ i, S i) → (∀ i, S i) → Prop)
(allow : (i : ι) → (∀ i, S i) → S i → Prop)
```

**Theorem 6.18 (hierarchy closure, soundness).** A node of sound protocols is a sound protocol. Its ballot of advance is a proof of the whole nested AND below it, and every commit inside it is a move of one child: a tick the node allowed, or a commit inside the child.

```
theorem Proto.node_sound (hP : ∀ i, (P i).Sound) :
(Proto.node P own pred allow).Sound
```

Theorem 6.19 (hierarchy closure, completeness). If every party is in the list all, a node of complete protocols is complete: every tick of the nested AND can be proposed from below.

```
theorem Proto.node_complete (all : List ι) (hall : ∀ i, i ∈ all)
(hPc : ∀ i, (P i).Complete) :
(Proto.node P own pred allow).Complete
```

**Theorem 6.20 (an allowed tick of one child can be committed alone).** If child i agrees to the tick to t and the node allows it, the node can reach a configuration from which child i is committed alone.

```
theorem Proto.node_delegate (i : ι) (hPc : (P i).Complete)
{g : ∀ i, S i} {c : (Proto.node P own pred allow).Cfg}
(hr : (Proto.node P own pred allow).Reach g c) (t : S i)
(hv : (P i).obj.vote (g i) t) (hg : allow i g t) :
∃ c1 c' : (Proto.node P own pred allow).Cfg,
(Proto.node P own pred allow).Steps g c c1 ∧
(Proto.node P own pred allow).istep g c1 (upd g i t) c'
```

### *The root*

Standing variables:

```
variable {S : Type}
```

**Definition 6.21 (Root).** The root drives a protocol. Root P g0 g c n: starting from committed state g0, the committed state is now g, the protocol is in configuration c, and the root's clock has ticked n times. commit is a tick of the root: the whole tentative state becomes history in one step. inner is a commit inside the tree: history changes and the root's clock does not tick.

```
inductive Root (P : Proto S) (g0 : S) : S → P.Cfg → Nat → Prop where
| start : Root P g0 g0 (P.idle g0) 0
| step {g : S} {c c' : P.Cfg} {n : Nat} :
Root P g0 g c n → P.step g c c' → Root P g0 g c' n
| commit {g : S} {c : P.Cfg} {n : Nat} :
Root P g0 g c n → P.ballot c = Ballot.advance →
Root P g0 (P.tent c) (P.idle (P.tent c)) (n + 1)
| abort {g : S} {c : P.Cfg} {n : Nat} :
Root P g0 g c n → P.ballot c = Ballot.retreat →
Root P g0 g (P.idle g) n
| inner {g g' : S} {c c' : P.Cfg} {n : Nat} :
Root P g0 g c n → P.istep g c g' c' → Root P g0 g' c' n
```

**Theorem 6.22 (the hierarchy refines the nested AND).** The committed history of a sound protocol is a history of its object: each root commit is a tick, each commit inside is an inner move, and the logical time of the history is the number of root commits.

```
theorem Root.history {P : Proto S} (hP : P.Sound) {g0 g : S} {c : P.Cfg}
{n : Nat} (r : Root P g0 g c n) :
∃ h : ORun P.obj g0 g, h.time = n
```

**Theorem 6.23 (the committed state is always inside the invariant).**

```
theorem Root.stat {P : Proto S} (hP : P.Sound) (hf : P.obj.Faithful)
{g0 g : S} {c : P.Cfg} {n : Nat} (r : Root P g0 g c n)
(h0 : P.obj.stat g0) : P.obj.stat g
```

**Theorem 6.24 (every tick of the nested AND is reachable).** For a complete protocol, from any point of a root run, a tick the object agrees to can be committed by the root.

```
theorem Root.complete {P : Proto S} (hPc : P.Complete) {g0 g g' : S}
{c : P.Cfg} {n : Nat} (r : Root P g0 g c n) (hv : P.obj.vote g g') :
Root P g0 g' (P.idle g') (n + 1)
```

**Theorem 6.25 (abort restores exactly).** When a sound protocol reports retreat, its tentative state is the committed state.

```
theorem Root.abort_restores {P : Proto S} (hP : P.Sound) {g0 g : S}
{c : P.Cfg} {n : Nat} (r : Root P g0 g c n)
(hb : P.ballot c = Ballot.retreat) : P.tent c = g
```

Standing variables:

```
variable {ι : Type} [DecidableEq ι] {κ : ι → Type} [∀ i, DecidableEq (κ i)]
{T : (i : ι) → κ i → Type}
```

**Definition 6.26.** Leaves under inner clocks under one outer clock.

```
def tree (o : ∀ i j, Obj (T i j))
(own₁ : ∀ i, (∀ j, T i j) → Prop)
(pred₁ : ∀ i, (∀ j, T i j) → (∀ j, T i j) → Prop)
(allow₁ : ∀ i, (j : κ i) → (∀ j, T i j) → T i j → Prop)
(own₀ : (∀ i j, T i j) → Prop)
(pred₀ : (∀ i j, T i j) → (∀ i j, T i j) → Prop)
(allow₀ : (i : ι) → (∀ i j, T i j) → (∀ j, T i j) → Prop) :
Proto (∀ i j, T i j) :=
Proto.node
(fun i => Proto.node (fun j => Proto.leaf (o i j))
(own₁ i) (pred₁ i) (allow₁ i))
own₀ pred₀ allow₀
```

**Theorem 6.27 (three levels).** The committed history of the three-level hierarchy is a history of the outer clock. Each commit at the root is one tick in which every leaf votes, every inner clock's predicates hold and the outer clock's predicates hold; the number of such ticks is the number of root commits.

```
theorem tree_history (o : ∀ i j, Obj (T i j))
(own₁ : ∀ i, (∀ j, T i j) → Prop)
(pred₁ : ∀ i, (∀ j, T i j) → (∀ j, T i j) → Prop)
(allow₁ : ∀ i, (j : κ i) → (∀ j, T i j) → T i j → Prop)
(own₀ : (∀ i j, T i j) → Prop)
(pred₀ : (∀ i j, T i j) → (∀ i j, T i j) → Prop)
(allow₀ : (i : ι) → (∀ i j, T i j) → (∀ j, T i j) → Prop)
{g0 g : ∀ i j, T i j}
{c : (tree o own₁ pred₁ allow₁ own₀ pred₀ allow₀).Cfg} {n : Nat}
(r : Root (tree o own₁ pred₁ allow₁ own₀ pred₀ allow₀) g0 g c n) :
∃ h : ORun (tree o own₁ pred₁ allow₁ own₀ pred₀ allow₀).obj g0 g,
h.time = n
```

## 7 Iteration on rejection

When a parent rejects a choice, the child offers another. This is the same pattern at every level: a node offers the combinations of its children's offers that its own predicates accept.

**Definition 7.1 (Chooser).** The choices an object can offer from committed state g: exactly the ticks it agrees to.

```
structure Chooser {S : Type} (o : Obj S) : Type where
offers : S → List S
sound : ∀ g t, t ∈ offers g → o.vote g t
complete : ∀ g t, o.vote g t → t ∈ offers g
```

**Definition 7.2 (pick).** The first choice not rejected, if any.

```
def Chooser.pick {S : Type} {o : Obj S} (ch : Chooser o) (g : S) : Option S :=
(ch.offers g).head?
```

**Theorem 7.3 (iteration ends in a tick).** If iteration returns a choice, the object agrees to that tick.

```
theorem Chooser.pick_sound {S : Type} {o : Obj S} (ch : Chooser o) {g t :
S}
(h : ch.pick g = some t) : o.vote g t
```

**Theorem 7.4 (iteration fails only when no tick exists).** If every choice has been rejected, the object agrees to no tick from g.

```
theorem Chooser.pick_none {S : Type} {o : Obj S} (ch : Chooser o) {g : S}
(h : ch.pick g = none) : ∀ t, ¬ o.vote g t
```

Standing variables:

```
variable {ι : Type} [DecidableEq ι] {S : ι → Type}
```

**Definition 7.5 (combos).** The combinations of the children's offers: for each party in the list, each of its offers in turn.

```
def combos (ch : ∀ i, List (S i)) (g : ∀ i, S i) : List ι → List (∀ i, S i)
| [] => [g]
| i :: l => (ch i).flatMap (fun s => (combos ch g l).map (fun f => upd f i s))
```

Standing variables:

```
variable {ι : Type} [DecidableEq ι] {S : ι → Type}
variable (objs : ∀ i, Obj (S i)) (own : (∀ i, S i) → Prop)
(pred : (∀ i, S i) → (∀ i, S i) → Prop)
(allow : (i : ι) → (∀ i, S i) → S i → Prop)
[∀ g t, Decidable (pred g t)]
```

**Definition 7.6 (a node of choosers is a chooser).** The node offers the combinations of its children's offers that its own predicates accept. Every other combination is a rejected choice.

```
def Chooser.node (all : List ι) (hall : ∀ i, i ∈ all)
(ch : ∀ i, Chooser (objs i)) :
Chooser (Clock.toObj ⟨objs, own, pred, allow⟩) where
offers g :=
(combos (fun i => (ch i).offers (g i)) g all).filter
(fun t => decide (pred g t))
sound := _ – proved in the Lean file
complete := _ – proved in the Lean file
```

Standing variables:

```
variable {ι : Type} [DecidableEq ι] {S : ι → Type}
```

**Definition 7.7.** The product of the lengths of the children's offers.

```
def prodLen (ch : ∀ i, List (S i)) : List ι → Nat
| [] => 1
| i :: l => (ch i).length * prodLen ch l
```

Standing variables:

```
variable {ι : Type} [DecidableEq ι] {S : ι → Type}
variable (objs : ∀ i, Obj (S i)) (own : (∀ i, S i) → Prop)
(pred : (∀ i, S i) → (∀ i, S i) → Prop)
(allow : (i : ι) → (∀ i, S i) → S i → Prop)
[∀ g t, Decidable (pred g t)]
```

**Theorem 7.8 (pruning is hierarchical).** A node examines at most the product of the numbers of choices its children offer. A choice a child has rejected is never seen by the parent.

```
theorem Chooser.node_offers_le (all : List ι) (hall : ∀ i, i ∈ all)
(ch : ∀ i, Chooser (objs i)) (g : ∀ i, S i) :
((Chooser.node objs own pred allow all hall ch).offers g).length ≤
prodLen (fun i => (ch i).offers (g i)) all
```

**Theorem 7.9 (a child with no choice makes the parent retreat).** If some child has no offer, the node has no offer.

```
theorem Chooser.node_no_offer (all : List ι) (hall : ∀ i, i ∈ all)
(ch : ∀ i, Chooser (objs i)) (g : ∀ i, S i) (i : ι)
(hi : (ch i).offers (g i) = []) :
(Chooser.node objs own pred allow all hall ch).pick g = none
```

Standing variables:

```
variable {ι : Type} [DecidableEq ι] {S : ι → Type}
(P : ∀ i, Proto (S i)) (own : (∀ i, S i) → Prop)
(pred : (∀ i, S i) → (∀ i, S i) → Prop)
(allow : (i : ι) → (∀ i, S i) → S i → Prop)
```

**Theorem 7.10 (a rejected choice does not percolate).** While the node's predicates fail on the tentative state, the node does not report advance, whatever its children report.

```
theorem Proto.node_rejected {g : ∀ i, S i}
{c : (Proto.node P own pred allow).Cfg}
(hr : (Proto.node P own pred allow).Reach g c)
(hrej : ¬ pred g (fun i => (P i).tent (c.1 i))) :
c.2 ≠ Ballot.advance
```

**Theorem 7.11 (the child must iterate).** If the parent's predicates reject the tentative state t, any later configuration in which the node reports advance has a different tentative state.

```
theorem Proto.node_must_iterate {g : ∀ i, S i}
{c : (Proto.node P own pred allow).Cfg}
(hr : (Proto.node P own pred allow).Reach g c)
(hb : c.2 = Ballot.advance) (t : ∀ i, S i) (hrej : ¬ pred g t) :
(fun i => (P i).tent (c.1 i)) ≠ t
```

Standing variables:

```
variable {S : Type}
```

**Theorem 7.12 (iteration drives the root).** If iteration at the root returns a choice, the root of a complete protocol commits it: one more tick of the nested clock.

```
theorem Root.iterate {P : Proto S} (hPc : P.Complete) (ch : Chooser P.obj)
{g0 g t : S} {c : P.Cfg} {n : Nat} (r : Root P g0 g c n)
(h : ch.pick g = some t) : Root P g0 t (P.idle t) (n + 1)
```

**Theorem 7.13 (the root retreats only when the clock cannot tick).** If iteration at the root returns nothing, no tick of the object exists from the committed state.

```
theorem Root.no_choice {P : Proto S} (ch : Chooser P.obj) {g : S}
(h : ch.pick g = none) : ∀ t, ¬ P.obj.vote g t
```

### 8 Many clocks

Every clock of the hierarchy has its own logical time and its own history. A child's time is at least its parent's.

Standing variables:

```
variable {ι : Type} [DecidableEq ι] {S : ι → Type}
```

**Theorem 8.1.** A history of a clock, as an object, is a history of the clock in which each inner move is a tick or a move of one named child. The logical time is the same.

```
theorem ORun.toCRun {C : Clock S} {g0 g : ∀ i, S i}
(h : ORun C.toObj g0 g) : ∃ r : CRun C g0 g, r.time = h.time
```

Standing variables:

```
variable {ι : Type} [DecidableEq ι] {S : ι → Type}
variable (P : ∀ i, Proto (S i)) (own : (∀ i, S i) → Prop)
(pred : (∀ i, S i) → (∀ i, S i) → Prop)
(allow : (i : ι) → (∀ i, S i) → S i → Prop)
```

**Theorem 8.2 (every clock has its own time).** In a run of the hierarchy, the committed history is a history of the root clock whose logical time is the number of root commits. Every child's logical time is at least that, and equal to it for a child the root never allows to advance alone.

```
theorem Root.clock_history (hP : ∀ i, (P i).Sound)
{g0 g : ∀ i, S i} {c : (Proto.node P own pred allow).Cfg} {n : Nat}
(r : Root (Proto.node P own pred allow) g0 g c n) :
∃ h : CRun ⟨fun i => (P i).obj, own, pred, allow⟩ g0 g,
h.time = n ∧ (∀ i, n ≤ h.childTime i) ∧
∀ i, (∀ g s', ¬ allow i g s') → h.childTime i = n
```

**Theorem 8.3 (every clock is contained in its own specification, at its own time).** Let obs map the root clock to a specification ANext, AHolds (Definition 5.2). In a run of the hierarchy, the history sampled at the root's ticks is a safe trace of that specification, and its logical time is the number of root commits, whatever the clocks below did on their own.

```
theorem Root.clock_contained {A : Type} (hP : ∀ i, (P i).Sound)
(hobj : ∀ i, (P i).obj.Faithful)
(hC : (⟨fun i => (P i).obj, own, pred, allow⟩ : Clock S).Sound)
{obs : (∀ i, S i) → A} {ANext : A → A → Prop} {AHolds : A → Prop}
(hA : (⟨fun i => (P i).obj, own, pred, allow⟩ : Clock S).Abstraction
obs ANext AHolds)
{g0 g : ∀ i, S i} {c : (Proto.node P own pred allow).Cfg} {n : Nat}
(r : Root (Proto.node P own pred allow) g0 g c n)
(h0 : (Clock.toObj ⟨fun i => (P i).obj, own, pred, allow⟩).stat g0) :
∃ h : CRun ⟨fun i => (P i).obj, own, pred, allow⟩ g0 g,
L_max ANext AHolds (h.sample obs) ∧ (h.sample obs).time = n
```

**Theorem 8.4 (a child's clock can tick while its parent's does not).** If child i agrees to the tick to t and its parent, in its current state, allows it, the hierarchy can commit it: the committed state changes in component i only, and the parent's logical time is unchanged.

```
theorem Root.delegate (i : ι) (hPc : (P i).Complete)
{g0 g : ∀ i, S i} {c : (Proto.node P own pred allow).Cfg} {n : Nat}
(r : Root (Proto.node P own pred allow) g0 g c n) (t : S i)
(hv : (P i).obj.vote (g i) t) (hg : allow i g t) :
∃ c' : (Proto.node P own pred allow).Cfg,
Root (Proto.node P own pred allow) g0 (upd g i t) c' n
```

Standing variables:

```
variable {ι : Type} [DecidableEq ι] {S : ι → Type}
```

**Theorem 8.5 (each clock has its own history).** A history of a clock, projected to child i, is a history of that child's object. Its logical time is the child's time (Definition 4.13).

```
theorem CRun.child {C : Clock S} (i : ι) {g h : ∀ i, S i} (r : CRun C g h) :
∃ o : ORun (C.obj i) (g i) (h i), o.time = r.childTime i
```

Standing variables:

```
variable {ι : Type} [DecidableEq ι] {κ : ι → Type} [∀ i, DecidableEq (κ i)]
{T : (i : ι) → κ i → Type} {A : Type}
```

**Theorem 8.6 (containment at every level, each at its own time).** Let the children of a clock be clocks D i. In any history of the parent, child i has a history of its own. Its logical time is the child's time. Sampled at the child's ticks it is a safe trace of the child's specification. If the child never allows its own child j to advance alone, the time of j in that history is the child's time.

```
theorem CRun.child_contained (D : ∀ i, Clock (T i))
(own : (∀ i j, T i j) → Prop)
(pred : (∀ i j, T i j) → (∀ i j, T i j) → Prop)
(allow : (i : ι) → (∀ i j, T i j) → (∀ j, T i j) → Prop)
(i : ι) (hobj : ∀ j, ((D i).obj j).Faithful) (hD : (D i).Sound)
{obs : (∀ j, T i j) → A} {ANext : A → A → Prop} {AHolds : A → Prop}
(hA : (D i).Abstraction obs ANext AHolds)
{g h : ∀ i j, T i j}
(r : CRun (S := fun i => ∀ j, T i j)
⟨fun i => (D i).toObj, own, pred, allow⟩ g h)
(hg : (D i).toObj.stat (g i)) :
∃ d : CRun (D i) (g i) (h i),
d.time = r.childTime i ∧
L_max ANext AHolds (d.sample obs) ∧
(d.sample obs).time = r.childTime i ∧
∀ j, (∀ s t, ¬ (D i).allow j s t) → d.childTime j = r.childTime i
```

## 9 The parent programs the clock

In Definition 4.5 the permission for a child to advance alone is read from the parent's whole state. This section treats the case in which the parent holds a program: a part of its state that no move inside the clock can change. Then autonomy is given, and taken back, only by a tick of the parent, and the decision stands until the parent's next tick.

Standing variables:

```
variable {ι : Type} [DecidableEq ι] {S : ι → Type} {R : Type}
```

**Definition 9.1 (Program).** prog reads a program out of the state of the clock. perm r i s s' says that program r permits child i to tick alone from s to s'. The clock is *programmed* by prog and perm when (fixed) no move inside the clock changes the program and (reads) the clock allows a child to advance alone only as the current program permits.

```
structure Clock.Program (C : Clock S) (prog : (∀ i, S i) → R)
(perm : R → (i : ι) → S i → S i → Prop) : Prop where
fixed : ∀ g g', C.toObj.inner g g' → prog g' = prog g
reads : ∀ i g s', C.allow i g s' → perm (prog g) i (g i) s'
```

**Theorem 9.2 (only the parent's tick changes the program).** In a history of a programmed clock with no tick of the clock, the program at the end is the program at the start.

```
theorem CRun.program_fixed {C : Clock S} {prog : (∀ i, S i) → R}
{perm : R → (i : ι) → S i → S i → Prop} (hp : C.Program prog perm)
{g h : ∀ i, S i} (r : CRun C g h) : r.time = 0 → prog h = prog g
```

**Theorem 9.3 (a decision of the parent stands until its next tick).** If the program at g permits child i no tick alone, then in every history from g with no tick of the clock, child i does not tick: its clock waits for its parent's.

```
theorem CRun.locked {C : Clock S} {prog : (∀ i, S i) → R}
{perm : R → (i : ι) → S i → S i → Prop} (hp : C.Program prog perm)
(i : ι) {g h : ∀ i, S i} (r : CRun C g h)
(hlock : ∀ s s', ¬ perm (prog g) i s s') :
r.time = 0 → r.childTime i = 0
```

**Theorem 9.4 (autonomy, once given, can be used).** If the program at the end of a history permits child i the tick to t, the clock allows what its program permits, and the child agrees, the history extends by that tick of the child alone. The clock's time is unchanged and the child's time is one more.

```
theorem CRun.autonomous {C : Clock S} {prog : (∀ i, S i) → R}
{perm : R → (i : ι) → S i → S i → Prop}
(hallow : ∀ i g s', perm (prog g) i (g i) s' → C.allow i g s')
(i : ι) {g h : ∀ i, S i} (r : CRun C g h) (t : S i)
(hv : (C.obj i).vote (h i) t) (hperm : perm (prog h) i (h i) t) :
∃ r' : CRun C g (upd h i t),
r'.time = r.time ∧ r'.childTime i = r.childTime i + 1
```

### *A clock with a program register*

Standing variables:

```
variable {ι : Type} {S : ι → Type} {R : Type}
```

**Definition 9.5 (PState).** The parties of a clock with a register: the register, at none, and the children.

```
def PState (R : Type) (S : ι → Type) : Option ι → Type
| none => R
| some i => S i
```

**Definition 9.6 (register).** An object that holds a value. It has no invariant, it makes no move of its own and it does not object to any tick.

```
def register (R : Type) : Obj R where
stat _ := True
diff _ _ := True
vote _ _ := True
inner _ _ := False
```

**Definition 9.7.** The objects of a clock with a register.

```
def regObj (R : Type) (obj : ∀ i, Obj (S i)) : (o : Option ι) → Obj (PState R S o)
| none => register R
| some i => obj i
```

**Definition 9.8.** What the register permits, as a relation on every party. Nothing moves the register alone.

```
def regPerm (perm : R → (i : ι) → S i → S i → Prop) :
R → (o : Option ι) → PState R S o → PState R S o → Prop
| _, none, _, _ => False
| r, some i, s, s' => perm r i s s'
```

**Definition 9.9 (programmed clock).** A clock whose program is the value of its register. Child i may tick alone from s to s' exactly when the register's value r has perm r i s s'. own and pred are the clock's predicates; pred decides, among other things, how the clock's own tick may rewrite the register.

```
def Clock.programmed (obj : ∀ i, Obj (S i))
(own : (∀ o, PState R S o) → Prop)
(pred : (∀ o, PState R S o) → (∀ o, PState R S o) → Prop)
(perm : R → (i : ι) → S i → S i → Prop) : Clock (PState R S) where
obj := regObj R obj
own := own
pred := pred
allow o g s' := regPerm perm (g none) o (g o) s'
```

**Theorem 9.10.** A clock with a register is programmed by it: only the clock's own tick rewrites the register, and a child advances alone only as the register permits.

```
theorem Clock.programmed_program (obj : ∀ i, Obj (S i))
(own : (∀ o, PState R S o) → Prop)
(pred : (∀ o, PState R S o) → (∀ o, PState R S o) → Prop)
(perm : R → (i : ι) → S i → S i → Prop) :
(Clock.programmed obj own pred perm).Program
(fun g => g none) (regPerm perm)
```

### *Only the immediate parent is asked*

Standing variables:

```
variable {ι : Type} [DecidableEq ι] {κ : ι → Type} [∀ i, DecidableEq (κ i)]
{T : (i : ι) → κ i → Type}
```

**Theorem 9.11 (only the immediate parent is asked).** In the three-level hierarchy of Definition 6.26, let leaf j of inner clock i agree to the tick to t, and let inner clock i, its immediate parent, allow it. Then the hierarchy commits that tick alone, whatever the outer clock allows: $\text{allow}_0$ is arbitrary and may be empty. The root's logical time is unchanged.

```
theorem tree_delegate (o : ∀ i j, Obj (T i j))
(own₁ : ∀ i, (∀ j, T i j) → Prop)
(pred₁ : ∀ i, (∀ j, T i j) → (∀ j, T i j) → Prop)
(allow₁ : ∀ i, (j : κ i) → (∀ j, T i j) → T i j → Prop)
(own₀ : (∀ i j, T i j) → Prop)
(pred₀ : (∀ i j, T i j) → (∀ i j, T i j) → Prop)
(allow₀ : (i : ι) → (∀ i j, T i j) → (∀ j, T i j) → Prop)
{g0 g : ∀ i j, T i j}
{c : (tree o own₁ pred₁ allow₁ own₀ pred₀ allow₀).Cfg} {n : Nat}
(r : Root (tree o own₁ pred₁ allow₁ own₀ pred₀ allow₀) g0 g c n)
(i : ι) (j : κ i) (t : T i j)
(hv : (o i j).vote (g i j) t) (ha : allow₁ i j (g i) t) :
∃ c' : (tree o own₁ pred₁ allow₁ own₀ pred₀ allow₀).Cfg,
Root (tree o own₁ pred₁ allow₁ own₀ pred₀ allow₀) g0
(upd g i (upd (g i) j t)) c' n
```

### *Example: autonomy given and taken back*

**Definition 9.12.** One counter under a parent whose program is one bit. The bit says whether the counter may tick alone. The parent's tick may set the bit either way.

```
– in namespace ProgramExample
def boss : Clock (PState Bool (fun _ : Unit => Nat)) :=
Clock.programmed (fun _ => counter) (fun _ => True) (fun _ _ => True)
(fun r _ _ _ => r = true)
```

**Definition 9.13.** The state with bit b and counter n.

```
– in namespace ProgramExample
def st (b : Bool) (n : Nat) : ∀ o, PState Bool (fun _ : Unit => Nat) o
| none => b
| some _ => n
```

**Example 9.14.** While the bit is off, the counter does not tick between two ticks of the parent.

```
– in namespace ProgramExample
theorem boss_locked (n : Nat) {h : ∀ o, PState Bool (fun _ : Unit => Nat) o}
(r : CRun boss (st false n) h) : r.time = 0 → r.childTime (some ()) = 0
```

**Example 9.15.** A history of four moves. The parent ticks and turns the bit on. The counter ticks alone twice. The parent ticks and turns the bit off.

```
– in namespace ProgramExample
def session : CRun boss (st false 0) (st false 4) – the term is in the Lean file
```

**Example 9.16.** In that history the parent's time is 2 and the counter's time is 4.

```
– in namespace ProgramExample
theorem session_times :
session.time = 2 ∧ session.childTime (some ()) = 4
```

## 10 L.0 programs

This section checks a selected core of the published L.0 syntax ([23], Sections IV and VI). now cause then { formula } is cause → next formula, and whenever cause then { formula } is always (cause → next formula). A cause is a current-state predicate. An effect is a predicate on the next state that may read previous values. The due effects determine the current obligation; the next state must satisfy it, and formulas triggered there become due at the following step. The object and nested-clock theorems check this rule at the governing clock's ticks. The original term "conditional invariant" below refers to this cause and effect rule. The core omits until and does not formalize the full IC* activation-and-termination mechanism.

Standing variables:

```
variable {S : Type}
```

**Definition 10.1 (L.0 formula).** tt: true. eff e: an effect predicate, e s s' where s' is the state being constructed and s the previous one. and: conjunction. now c f: c → next f. whenever c f: always (c → next f).

```
inductive L0F (S : Type) : Type where
| tt : L0F S
| eff (e : S → S → Prop) : L0F S
| and (f g : L0F S) : L0F S
| now (c : S → Bool) (f : L0F S) : L0F S
| whenever (c : S → Bool) (f : L0F S) : L0F S
```

**Definition 10.2 (current obligation).** What formula f requires of the state s' being constructed, the previous state being s.

```
def L0F.ob : L0F S → S → S → Prop
| .tt, _, _ => True
| .eff e, s, s' => e s s'
| .and f g, s, s' => f.ob s s' ∧ g.ob s s'
| .now _ _, _, _ => True
| .whenever _ _, _, _ => True
```

**Definition 10.3 (future obligation).** The formula that is due at the following step, once s' has been constructed: each cause is evaluated on s'.

```
def L0F.nx : L0F S → S → L0F S
| .tt, _ => .tt
| .eff _, _ => .tt
| .and f g, s' => .and (f.nx s') (g.nx s')
| .now c f, s' => if c s' = true then f else .tt
| .whenever c f, s' =>
if c s' = true then .and f (.whenever c f) else .whenever c f
```

**Definition 10.4 (an L.0 program as an object).** The state is the data state and the formula that is due. The object agrees to a tick exactly when the new data state meets the current obligation and the static invariant, and the formula is rewritten to the future obligation. It makes no move between ticks.

```
def Obj.l0 (inv : S → Prop) : Obj (S × L0F S) where
stat x := inv x.1
diff x x' := x.2.ob x.1 x'.1
vote x x' := x.2.ob x.1 x'.1 ∧ x'.2 = x.2.nx x'.1 ∧ inv x'.1
inner _ _ := False
```

**Theorem 10.5.** An L.0 program is a faithful object.

```
theorem Obj.l0_faithful (inv : S → Prop) : (Obj.l0 inv).Faithful
```

**Definition 10.6 (Always).** The conditional invariant whenever c then e is active in a formula.

```
inductive L0F.Always (c : S → Bool) (e : S → S → Prop) : L0F S → Prop where
| here : L0F.Always c e (.whenever c (.eff e))
| left {f g : L0F S} : L0F.Always c e f → L0F.Always c e (.and f g)
| right {f g : L0F S} : L0F.Always c e g → L0F.Always c e (.and f g)
```

**Theorem 10.7 (whenever cause then effect).** Let whenever c then e be active in an L.0 program. If the cause holds after one tick of the program, the effect holds across its next tick. The conditional invariant is still active afterwards.

```
theorem Obj.l0_whenever (inv : S → Prop) {c : S → Bool} {e : S → S → Prop}
{x x1 x2 : S × L0F S} (ha : x.2.Always c e)
(h1 : (Obj.l0 inv).vote x x1) (h2 : (Obj.l0 inv).vote x1 x2) :
x2.2.Always c e ∧ (c x1.1 = true → e x1.1 x2.1)
```

Standing variables:

```
variable {ι : Type} {S : ι → Type}
```

**Definition 10.8.** The children's part of the state of a clock with a register.

```
def kidsOf {R : Type} (g : ∀ o, PState R S o) : ∀ i, S i := fun i => g (some i)
```

**Definition 10.9 (L.0 clock).** A clock whose register holds an L.0 formula over the joint state of its children. The clock ticks when every child agrees, the new joint state meets the current obligation of the formula, and the register is rewritten to the future obligation. perm reads from the formula which child may advance alone.

```
def Clock.l0 (obj : ∀ i, Obj (S i)) (own : (∀ i, S i) → Prop)
(perm : L0F (∀ i, S i) → (i : ι) → S i → S i → Prop) :
Clock (PState (L0F (∀ i, S i)) S) :=
Clock.programmed obj (fun g => own (kidsOf g))
(fun g g' => L0F.ob (g none) (kidsOf g) (kidsOf g') ∧
g' none = L0F.nx (g none) (kidsOf g'))
perm
```

**Theorem 10.10 (conditional invariants are aligned with the clock).** Let whenever c then e be active in the program of a clock. Let the clock tick from g to g1 with the cause true on g1. Let the children then make any moves of their own, with no tick of the clock, reaching g1'. Then across the clock's next tick, from g1' to g2, the effect holds. The cause was read at the clock's tick; the effect is due at the clock's next tick; the children's own ticks in between did not change the program, did not trigger the invariant and did not discharge it. The effect relates the immediate pre-state g1′ to g2, rather than necessarily the earlier parent sample g1; comparing parent samples requires a snapshot or a suitable stability hypothesis.

```
theorem Clock.l0_whenever (obj : ∀ i, Obj (S i)) (own : (∀ i, S i) → Prop)
(perm : L0F (∀ i, S i) → (i : ι) → S i → S i → Prop)
{c : (∀ i, S i) → Bool} {e : (∀ i, S i) → (∀ i, S i) → Prop}
{g g1 g1' g2 : ∀ o, PState (L0F (∀ i, S i)) S o}
(ha : L0F.Always c e (g none))
(h1 : (Clock.l0 obj own perm).toObj.vote g g1)
(hc : c (kidsOf g1) = true)
(r : CRun (Clock.l0 obj own perm) g1 g1') (hr : r.time = 0)
(h2 : (Clock.l0 obj own perm).toObj.vote g1' g2) :
e (kidsOf g1') (kidsOf g2) ∧ L0F.Always c e (g2 none)
```

### *Example: the clock of the L.0 paper*

**Definition 10.11.** whenever true then clock = clock' + 1.

```
– in namespace L0Example
def clockF : L0F Nat :=
.whenever (fun _ => true) (.eff (fun s s' => s' = s + 1))
```

**Example 10.12.** After its first tick, the program adds one at every tick.

```
– in namespace L0Example
theorem clock_counts {x x1 x2 : Nat × L0F Nat} (hx : x.2 = clockF)
(h1 : (Obj.l0 (fun _ => True)).vote x x1)
(h2 : (Obj.l0 (fun _ => True)).vote x1 x2) : x2.1 = x1.1 + 1
```

## 11 Compositional verification

A clock sees its children only through their specifications. An object refines a specification when its ticks are ticks of the specification and its inner moves are inner moves of the specification or change nothing the specification can see. Refinement is preserved by the clock constructor. So a clock is verified against the specifications of its children, never against their implementations.

Standing variables:

```
variable {S A B : Type}
```

**Definition 11.1 (Refines).** Object c refines object a under the map f. The invariant of c maps into the invariant of a. A tick of c is a tick of a. An inner move of c is an inner move of a or is invisible under f.

```
structure Obj.Refines (c : Obj S) (a : Obj A) (f : S → A) : Prop where
stat : ∀ s, c.stat s → a.stat (f s)
vote : ∀ s s', c.stat s → c.vote s s' → a.vote (f s) (f s')
inner : ∀ s s', c.stat s → c.inner s s' →
a.inner (f s) (f s') ∨ f s' = f s
```

**Theorem 11.2.** Refinement is reflexive.

```
theorem Obj.Refines.refl (o : Obj S) : o.Refines o id
```

**Theorem 11.3.** Refinement is transitive.

```
theorem Obj.Refines.trans {c : Obj S} {b : Obj A} {a : Obj B}
{f : S → A} {h : A → B} (h1 : c.Refines b f) (h2 : b.Refines a h) :
c.Refines a (fun s => h (f s))
```

**Definition 11.4 (Image).** History o of the specification is the image under f of history r of the implementation, step by step: a tick of r is a tick of o; an inner move of r is an inner move of o, or is invisible and leaves o as it is.

```
inductive ORun.Image {c : Obj S} {a : Obj A} (f : S → A) {g : S} :
{h : S} → ORun c g h → ORun a (f g) (f h) → Prop where
| start : ORun.Image f .start .start
| tick {h h' : S} {r : ORun c g h} {o : ORun a (f g) (f h)}
(hv : c.vote h h') (hv' : a.vote (f h) (f h')) :
ORun.Image f r o → ORun.Image f (.tick r hv) (.tick o hv')
| inner {h h' : S} {r : ORun c g h} {o : ORun a (f g) (f h)}
(hi : c.inner h h') (hi' : a.inner (f h) (f h')) :
ORun.Image f r o → ORun.Image f (.inner r hi) (.inner o hi')
| stutter {h h' : S} {r : ORun c g h} {o : ORun a (f g) (f h)}
(hi : c.inner h h') (e : f h = f h') :
ORun.Image f r o → ORun.Image f (.inner r hi) (e ▸ o)
```

**Theorem 11.5 (a history of the implementation maps to a history of the specification, step by step and at the same logical time).** Let c be faithful and refine a under f. Every history of c that starts inside the invariant has an image (Definition 11.4) that is a history of a with the same number of ticks.

```
theorem ORun.refine {c : Obj S} {a : Obj A} {f : S → A}
(hf : c.Faithful) (hr : c.Refines a f) {g h : S} (r : ORun c g h)
(hg : c.stat g) :
∃ o : ORun a (f g) (f h), ORun.Image f r o ∧ o.time = r.time
```

Standing variables:

```
variable {ι : Type} [DecidableEq ι] {S A : ι → Type} {B : Type}
```

**Definition 11.6 (lift).** The joint state as the parent sees it: each child's state under that child's map.

```
def lift (f : ∀ i, S i → A i) (g : ∀ i, S i) : ∀ i, A i := fun i => f i (g i)
```

**Definition 11.7 (Over).** Clock C, over implementations, sits over clock CA, over specifications, when each child of C refines the corresponding child of CA, the two clocks own the same predicate as seen through f, and C requires at least what CA requires for a tick and for a child to advance alone.

```
structure Clock.Over (C : Clock S) (CA : Clock A) (f : ∀ i, S i → A i) :

Prop where
obj : ∀ i, (C.obj i).Refines (CA.obj i) (f i)
own : ∀ g, C.own g ↔ CA.own (lift f g)
pred : ∀ g g', C.pred g g' → CA.pred (lift f g) (lift f g')
allow : ∀ i g s', C.allow i g s' → CA.allow i (lift f g) (f i s')
```

**Theorem 11.8 (refinement is preserved by the clock constructor).** If C sits over CA, the object C refines the object CA.

```
theorem Clock.Over.refines {C : Clock S} {CA : Clock A}
{f : ∀ i, S i → A i} (h : C.Over CA f) :
C.toObj.Refines CA.toObj (lift f)
```

**Theorem 11.9 (the check is made on the specifications).** If C sits over CA and CA is sound, C is sound. The obligation of Definition 4.8 is discharged on the state spaces of the children's specifications.

```
theorem Clock.Over.sound {C : Clock S} {CA : Clock A}
{f : ∀ i, S i → A i} (h : C.Over CA f) (hA : CA.Sound) : C.Sound
```

**Theorem 11.10 (the specification of the clock is established on the specifications of its children).** If C sits over CA and obs maps CA to a specification (Definition 5.2), then obs after lift f maps C to the same specification.

```
theorem Clock.Over.abstraction {C : Clock S} {CA : Clock A}
{f : ∀ i, S i → A i} (h : C.Over CA f)
{obs : (∀ i, A i) → B} {ANext : B → B → Prop} {AHolds : B → Prop}
(hA : CA.Abstraction obs ANext AHolds) :
C.Abstraction (fun g => obs (lift f g)) ANext AHolds
```

**Definition 11.11 (pull).** The clock CA, stated over specifications, placed over implementations c that are read through f.

```
def Clock.pull (CA : Clock A) (c : ∀ i, Obj (S i)) (f : ∀ i, S i → A i) :

Clock S where
obj := c
own g := CA.own (lift f g)
pred g g' := CA.pred (lift f g) (lift f g')
allow i g s' := CA.allow i (lift f g) (f i s')
```

**Theorem 11.12.** A clock placed over implementations that refine its children sits over itself.

```
theorem Clock.pull_over (CA : Clock A) (c : ∀ i, Obj (S i))
(f : ∀ i, S i → A i) (hc : ∀ i, (c i).Refines (CA.obj i) (f i)) :
(CA.pull c f).Over CA f
```

**Theorem 11.13 (verify a clock against the specifications of its children).** Let CA be a sound clock over specifications that refines its own specification b. Place it over any faithful implementations that refine those specifications. The result is a faithful object and refines b.

```
theorem Clock.verify (CA : Clock A) (c : ∀ i, Obj (S i))
(f : ∀ i, S i → A i) (hcf : ∀ i, (c i).Faithful)
(hc : ∀ i, (c i).Refines (CA.obj i) (f i)) (hA : CA.Sound)
{b : Obj B} {h : (∀ i, A i) → B} (hb : CA.toObj.Refines b h) :
(CA.pull c f).toObj.Faithful ∧
(CA.pull c f).toObj.Refines b (fun g => h (lift f g))
```

## 12 Hierarchies of any depth

A hierarchy is an object or a clock over hierarchies. A design is a hierarchy in which every clock carries its specification. The local obligations of a design mention, at each clock, only the specifications of that clock's children. If they hold, the whole hierarchy refines the specification at its root.

**Definition 12.1 (Sys).** A hierarchy of clocks of any depth and any branching: an object, or a clock over hierarchies.

```
inductive Sys : Type → Type 1 where
| leaf {S : Type} (o : Obj S) : Sys S
| node {ι : Type} [DecidableEq ι] {S : ι → Type}
(kids : ∀ i, Sys (S i)) (own : (∀ i, S i) → Prop)
(pred : (∀ i, S i) → (∀ i, S i) → Prop)
(allow : (i : ι) → (∀ i, S i) → S i → Prop) : Sys (∀ i, S i)
```

**Definition 12.2.** The cycle protocol of a hierarchy (Definitions 6.12, 6.17).

```
def Sys.proto : {S : Type} → Sys S → Proto S
| _, .leaf o => Proto.leaf o
| _, @Sys.node ι _ S kids own pred allow =>
Proto.node (ι := ι) (S := S) (fun i => (kids i).proto) own pred allow
```

**Definition 12.3.** The object of a hierarchy: the nested AND.

```
def Sys.obj {S : Type} (t : Sys S) : Obj S := t.proto.obj
```

**Theorem 12.4.** The protocol of every hierarchy is sound.

```
theorem Sys.proto_sound {S : Type} (t : Sys S) : t.proto.Sound
```

**Definition 12.5.** Every clock of the hierarchy has finitely many parties.

```
def Sys.Fin : {S : Type} → Sys S → Prop
| _, .leaf _ => True
| _, @Sys.node ι _ _ kids _ _ _ =>
(∃ all : List ι, ∀ i, i ∈ all) ∧ ∀ i, (kids i).Fin
```

**Theorem 12.6.** The protocol of every finitely branching hierarchy is complete.

```
theorem Sys.proto_complete {S : Type} (t : Sys S) :
t.Fin → t.proto.Complete
```

**Definition 12.7 (Design).** A hierarchy in which every clock carries its specification. A leaf is an object o, its specification a and the map f between them. A node has children; its predicates and its program are stated over the *specifications* of the children; b is the node's own specification and h the map to it.

```
inductive Design : Type → Type → Type 1 where
| leaf {S A : Type} (o : Obj S) (a : Obj A) (f : S → A) : Design S A
| node {ι : Type} [DecidableEq ι] {S A : ι → Type} {B : Type}
(kids : ∀ i, Design (S i) (A i)) (own : (∀ i, A i) → Prop)
(pred : (∀ i, A i) → (∀ i, A i) → Prop)
(allow : (i : ι) → (∀ i, A i) → A i → Prop)
(b : Obj B) (h : (∀ i, A i) → B) : Design (∀ i, S i) B
```

**Definition 12.8.** The specification at the root.

```
def Design.abs : {S A : Type} → Design S A → Obj A
| _, _, .leaf _ a _ => a
| _, _, @Design.node _ _ _ _ _ _ _ _ _ b _ => b
```

**Definition 12.9.** The map from the state of the whole hierarchy to the state of the root's specification.

```
def Design.map : {S A : Type} → Design S A → S → A
| _, _, .leaf _ _ f => f
| _, _, @Design.node _ _ _ _ _ kids _ _ _ _ h =>
fun g => h (fun i => (kids i).map (g i))
```

**Definition 12.10.** The hierarchy itself: every node's predicates and program read its children through their maps.

```
def Design.sys : {S A : Type} → Design S A → Sys S
| _, _, .leaf o _ _ => .leaf o
| _, _, @Design.node ι _ S _ _ kids own pred allow _ _ =>
Sys.node (ι := ι) (S := S) (fun i => (kids i).sys)
(fun g => own (fun i => (kids i).map (g i)))
(fun g g' => pred (fun i => (kids i).map (g i))
(fun i => (kids i).map (g' i)))
(fun i g s' => allow i (fun i => (kids i).map (g i))
((kids i).map s'))
```

**Definition 12.11 (Local).** The local obligations. At a leaf: the object is faithful and refines its specification. At a node: the clock over the specifications of the children is sound (Definition 4.8) and refines the node's specification. No obligation mentions the state of an implementation below the node's children.

```
def Design.Local : {S A : Type} → Design S A → Prop
| _, _, .leaf o a f => o.Faithful ∧ o.Refines a f
| _, _, @Design.node _ _ _ _ _ kids own pred allow b h =>
(∀ i, (kids i).Local) ∧
Clock.Sound ⟨fun i => (kids i).abs, own, pred, allow⟩ ∧
(Clock.toObj ⟨fun i => (kids i).abs, own, pred, allow⟩).Refines b h
```

**Theorem 12.12 (local obligations suffice, at any depth).** If every clock of a hierarchy meets its local obligations, the whole hierarchy is a faithful object and refines the specification at its root.

```
theorem Design.verified {S A : Type} (t : Design S A) :
t.Local → t.sys.obj.Faithful ∧ t.sys.obj.Refines t.abs t.map
```

**Theorem 12.13 (every commit of the hierarchy is a tick of the root's specification).** Let every clock meet its local obligations. In any run of the hierarchy's protocol from a state inside the invariant, the committed state stays inside the invariant of every clock. The committed history has an image, step by step (Definition 11.4), that is a history of the root's specification, and the logical time of the image is the number of root commits. This is a statement of safety: it does not say that the hierarchy can do everything its specification allows.

```
theorem Design.commits {S A : Type} (t : Design S A) (hl : t.Local)
{g0 g : S} {c : t.sys.proto.Cfg} {n : Nat}
(r : Root t.sys.proto g0 g c n) (h0 : t.sys.obj.stat g0) :
t.sys.obj.stat g ∧
∃ (o : ORun t.sys.obj g0 g) (o' : ORun t.abs (t.map g0) (t.map g)),
ORun.Image t.map o o' ∧ o'.time = n
```

**Theorem 12.14 (every level, at its own time).** Let a node and all clocks below it meet their local obligations. For every history of the node there is a history of the *specification* of child i, between the images of the child's states at the two ends, whose logical time is at least the node's. Since the child is again a hierarchy that meets its local obligations, the statement applies to it in turn, to any depth.

```
theorem Design.child_history {ι : Type} [DecidableEq ι] {S A : ι → Type}
{B : Type} (kids : ∀ i, Design (S i) (A i)) (own : (∀ i, A i) → Prop)
(pred : (∀ i, A i) → (∀ i, A i) → Prop)
(allow : (i : ι) → (∀ i, A i) → A i → Prop) (b : Obj B)
(h : (∀ i, A i) → B) (hk : ∀ i, (kids i).Local) (i : ι)
{g g' : ∀ i, S i}
(o : ORun (Design.node kids own pred allow b h).sys.obj g g')
(hg : (kids i).sys.obj.stat (g i)) :
∃ o' : ORun (kids i).abs ((kids i).map (g i)) ((kids i).map (g' i)),
o.time ≤ o'.time
```

**Theorem 12.15 (every tick of the nested AND can be committed, at any depth).** In a finitely branching hierarchy, from any point of a run, a tick the nested AND agrees to can be committed by the root.

```
theorem Sys.complete {S : Type} (t : Sys S) (hfin : t.Fin) {g0 g g' : S}
{c : t.proto.Cfg} {n : Nat} (r : Root t.proto g0 g c n)
(hv : t.obj.vote g g') : Root t.proto g0 g' (t.proto.idle g') (n + 1)
```

### *Example: one resource, a tree of any depth*

Every leaf is a worker that may hold a resource. Every clock owns the predicate "at most one of my two children holds it" and gives autonomy to the child that holds it. One check on a clock with four abstract states verifies the tree of every depth.

**Definition 12.16.** The specification of every subtree: one bit, "the resource is held here". The bit may change at a tick. Between ticks it may only be given up.

```
– in namespace MutexExample
def flag : Obj Bool where
stat _ := True
diff _ _ := True
vote _ _ := True
inner b b' := b = true ∧ b' = false
```

**Definition 12.17.** A worker: a count of its ticks and the bit.

```
– in namespace MutexExample
def worker : Obj (Nat × Bool) where
stat _ := True
diff _ _ := True
vote s s' := s'.1 = s.1 + 1
inner _ _ := False
```

**Definition 12.18.** The clock of every node, over the specifications of its two children. It owns "not both hold the resource". Its program: the child that holds the resource may advance alone.

```
– in namespace MutexExample
def nodeA : Clock (fun _ : Bool => Bool) where
obj _ := flag
own g := ¬ (g false = true ∧ g true = true)
pred _ g' := ¬ (g' false = true ∧ g' true = true)
allow i g _ := g i = true
```

**Definition 12.19.** The map from a node to its specification: one of the children holds.

```
– in namespace MutexExample
def join (g : Bool → Bool) : Bool := g false || g true
```

**Example 12.20.** The local check, first half: the node is sound.

```
– in namespace MutexExample
theorem nodeA_sound : nodeA.Sound
```

**Example 12.21.** The local check, second half: the node refines the one-bit specification.

```
– in namespace MutexExample
theorem nodeA_refines : nodeA.toObj.Refines flag join
```

**Definition 12.22.** The state of the tree of depth d.

```
– in namespace MutexExample
def St : Nat → Type
| 0 => Nat × Bool
| d + 1 => Bool → St d
```

**Definition 12.23.** The tree of depth d: 2 ^ d workers under 2 ^ d - 1 clocks.

```
– in namespace MutexExample
def tree : (d : Nat) → Design (St d) Bool
| 0 => .leaf worker flag Prod.snd
| d + 1 => .node (fun _ => tree d) nodeA.own nodeA.pred nodeA.allow flag join
```

**Example 12.24.** Every clock of the tree of every depth meets its local obligations. The proof uses 12.20 and 12.21 once per level and nothing else.

```
– in namespace MutexExample
theorem tree_local : ∀ d, (tree d).Local
```

**Definition 12.25.** The number of workers that hold the resource.

```
– in namespace MutexExample
def held : (d : Nat) → St d → Nat
| 0, s => if s.2 = true then 1 else 0
| d + 1, g => held d (g false) + held d (g true)
```

**Example 12.26.** Inside the invariant of the tree, the number of workers that hold the resource is 1 if the root's bit is set and 0 otherwise.

```
– in namespace MutexExample
theorem held_eq : ∀ d (g : St d), (tree d).sys.obj.stat g →
held d g = if (tree d).map g = true then 1 else 0
```

**Example 12.27 (mutual exclusion at every depth, for every interleaving).** In every history of the tree of depth d that starts inside the invariant, whatever ticks and moves of single clocks it is made of, at most one worker holds the resource.

```
– in namespace MutexExample
theorem no_two_holders (d : Nat) {g0 g : St d}
(o : ORun (tree d).sys.obj g0 g) (h0 : (tree d).sys.obj.stat g0) :
held d g ≤ 1
```

**Example 12.28.** The same for every run of the protocol of the tree.

```
– in namespace MutexExample
theorem mutex (d : Nat) {g0 g : St d} {c : (tree d).sys.proto.Cfg} {n : Nat}
(r : Root (tree d).sys.proto g0 g c n)
(h0 : (tree d).sys.obj.stat g0) : held d g ≤ 1
```

**Theorem 12.29.** Every clock of the tree has finitely many parties.

```
– in namespace MutexExample
theorem tree_fin : ∀ d, (tree d).sys.Fin
```

**Example 12.30 (the holder runs on its own clock).** In the tree of depth d + 1, if the subtree i holds the resource and agrees to a tick to t, the hierarchy commits that tick of the subtree alone. The root's logical time is unchanged and the root is the only clock asked.

```
– in namespace MutexExample
theorem holder_runs_alone (d : Nat) {g0 g : St (d + 1)}
{c : (tree (d + 1)).sys.proto.Cfg} {n : Nat}
(r : Root (tree (d + 1)).sys.proto g0 g c n) (i : Bool) (t : St d)
(hv : (tree d).sys.obj.vote (g i) t)
(hheld : (tree d).map (g i) = true) :
∃ c', Root (tree (d + 1)).sys.proto g0 (upd g i t) c' n
```

The Lean file continues the example at depth 2: a run with three time scales (root time 1, subtree time 2, worker time 3) and a move that the immediate parent refuses (MutexExample.depth2_run, MutexExample.x2_refused).

## 13 A hierarchy of AI agents

The clock model is applied to a team with opaque agent objects and an arbitrary supervisory vote relation H. The orchestrator guards joint commits and permitted autonomous commits against the safe predicate. Its desk records autonomy and the last signed product. Autonomy may be granted only at a team tick, and may decrease between ticks when H admits a withdrawal preserving the signed record. This permission discipline is not the fixed Program condition of Definition 9.1. The transition model assumes that covered external effects are mediated through the corresponding commit boundary.

The theorems about runs of the protocol assume nothing about the agents and nothing about the supervisor. They assume the architecture: committed state changes only through the protocol of Section 6, and the orchestrator evaluates the guardrail on the true product. The theorem about the semantics (Theorem 13.20) lets an agent move by itself between commits and then needs the sandbox assumption: such a move does not change the agent's product.

Standing variables:

```
variable {S : Type}
```

**Lemma 13.1.** A predicate kept by every commit of a protocol, at the root and inside, is true of every committed state of a run.

```
theorem Root.invariant {P : Proto S} (Inv : S → Prop)
(hadv : ∀ g c, P.Reach g c → Inv g → P.ballot c = Ballot.advance →
Inv (P.tent c))
(hin : ∀ g c g' c', P.Reach g c → Inv g → P.istep g c g' c' → Inv g')
{g0 g : S} {c : P.Cfg} {n : Nat} (r : Root P g0 g c n) (h0 : Inv g0) :
Inv g
```

Standing variables:

```
variable {ι : Type} [DecidableEq ι] {W : ι → Type}
```

Definition 13.2 (black box). The specification of a party treated as opaque, with product of type V: it agrees to any change of its product at a tick, and its product does not change between ticks.

```
def blackbox (V : Type) : Obj V where
stat _ := True
diff _ _ := True
vote _ _ := True
inner _ _ := False
```

**Theorem 13.3 (every sandboxed agent is a black box).** Any object, with any state and any behaviour, whose moves between ticks leave its product f unchanged, refines the black box.

```
theorem Obj.refines_blackbox {S V : Type} (c : Obj S) (f : S → V)
(hsand : ∀ s s', c.stat s → c.inner s s' → f s' = f s) :
c.Refines (blackbox V) f
```

**Definition 13.4 (opaque).** An agent given by two arbitrary relations: what it proposes at a tick and what it does in private.

```
def Obj.opaque {S : Type} (propose scratch : S → S → Prop) : Obj S where
stat _ := True
diff _ _ := True
vote := propose
inner := scratch
```

**Definition 13.5 (desk).** The state of the supervisor: which agents have autonomy, and the joint product it last signed.

```
structure Desk (W : ι → Type) : Type where
auto : ι → Bool
seen : ∀ i, W i
```

**Definition 13.6 (supervisor).** A party whose decisions are the arbitrary relation H: it agrees to move its desk from d to d' exactly when H d d'.

```
def supervisor (H : Desk W → Desk W → Prop) : Obj (Desk W) where
stat _ := True
diff _ _ := True
vote := H
inner _ _ := False
```

**Definition 13.7.** The parties of a team, as the orchestrator sees them: the supervisor and one black box per agent.

```
def teamObj (H : Desk W → Desk W → Prop) :
(o : Option ι) → Obj (PState (Desk W) W o)
| none => supervisor H
| some i => blackbox (W i)
```

Definition 13.8 (the permission relation of a team). The supervisor advances alone only to withdraw: its new desk gives autonomy to no agent that did not have it, and its record of what it signed is unchanged. Agent i advances alone when the desk gives it autonomy and the guardrail holds on the result.

```
def teamAllow (safe : (∀ i, W i) → Prop) :
(o : Option ι) → (∀ o, PState (Desk W) W o) → PState (Desk W) W o → Prop
| none, g, d' =>
(∀ i, Desk.auto d' i = true → Desk.auto (g none) i = true) ∧
Desk.seen d' = Desk.seen (g none)
| some i, g, s' => Desk.auto (g none) i = true ∧ safe (upd (kidsOf g) i s')
```

**Definition 13.9 (team).** The orchestrator's clock. It owns the guardrail safe. It ticks when every party agrees, the supervisor's desk records exactly the joint product being committed, and the guardrail holds on it.

```
def Clock.team (H : Desk W → Desk W → Prop) (safe : (∀ i, W i) → Prop) :
Clock (PState (Desk W) W) where
obj := teamObj H
own g := safe (kidsOf g)
pred _ g' := Desk.seen (g' none) = kidsOf g' ∧ safe (kidsOf g')
allow := teamAllow safe
```

**Theorem 13.10 (the local check).** The team is sound: every tick and every move of one party alone keeps the guardrail.

```
theorem Clock.team_sound (H : Desk W → Desk W → Prop)
(safe : (∀ i, W i) → Prop) : (Clock.team H safe).Sound
```

**Definition 13.11 (guard).** The specification of a team as seen from outside: its joint product satisfies the guardrail, at every tick and after every move inside.

```
def guard {V : Type} (safe : V → Prop) : Obj V where
stat := safe
diff _ _ := True
vote _ v' := safe v'
inner _ v' := safe v'
```

**Theorem 13.12.** The team refines its guardrail.

```
theorem Clock.team_refines (H : Desk W → Desk W → Prop)
(safe : (∀ i, W i) → Prop) :
(Clock.team H safe).toObj.Refines (guard safe) kidsOf
```

Theorem 13.13 (autonomy cannot increase between team ticks). In a history of the team with no tick, no agent gains autonomy, and an agent that has none at the start does not commit: its clock waits for the team's.

```
theorem Clock.team_grants (H : Desk W → Desk W → Prop)
(safe : (∀ i, W i) → Prop) {g h : ∀ o, PState (Desk W) W o}
(r : CRun (Clock.team H safe) g h) :
r.time = 0 →
(∀ i, Desk.auto (h none) i = true → Desk.auto (g none) i = true) ∧
∀ i, Desk.auto (g none) i = false → r.childTime (some i) = 0
```

**Theorem 13.14 (a team is again an agent).** Let pub be what a team publishes of its joint product, and let no move of one agent alone change what is published. Then the team refines the black box, and can be an agent of a larger team. The larger team sees the smaller one change only at the smaller one's own ticks.

```
theorem Clock.team_blackbox (H : Desk W → Desk W → Prop)
(safe : (∀ i, W i) → Prop) {V : Type} (pub : (∀ i, W i) → V)
(hpub : ∀ g i s', safe g → safe (upd g i s') → pub (upd g i s') = pub g) :
(Clock.team H safe).toObj.Refines (blackbox V)
(fun g => pub (kidsOf g))
```

### *Any agents, any supervisor*

Standing variables:

```
variable {ι : Type} [DecidableEq ι] {W : ι → Type}
variable {S : ι → Type}
```

**Definition 13.15.** The parties of a team with their implementations: the supervisor as it is, and for each agent an arbitrary object c i whose product is read by f i.

```
def teamKids (H : Desk W → Desk W → Prop) (c : ∀ i, Obj (S i))
(f : ∀ i, S i → W i) :
(o : Option ι) → Design (PState (Desk W) S o) (PState (Desk W) W o)
| none => .leaf (supervisor H) (supervisor H) id
| some i => .leaf (c i) (blackbox (W i)) (f i)
```

**Definition 13.16 (a team of arbitrary agents).** The hierarchy with the supervisor and the agents c i at the leaves and the orchestrator's clock above them. Its specification is the guardrail.

```
def teamDesign (H : Desk W → Desk W → Prop) (safe : (∀ i, W i) → Prop)
(c : ∀ i, Obj (S i)) (f : ∀ i, S i → W i) :
Design (∀ o, PState (Desk W) S o) (∀ i, W i) :=
Design.node (S := PState (Desk W) S) (A := PState (Desk W) W)
(teamKids H c f) (fun g => safe (kidsOf g))
(fun _ g' => Desk.seen (g' none) = kidsOf g' ∧ safe (kidsOf g'))
(teamAllow safe) (guard safe) kidsOf
```

**Definition 13.17.** The joint product of the agents in a state of the team.

```
def teamProduct (f : ∀ i, S i → W i) (g : ∀ o, PState (Desk W) S o) :
∀ i, W i := fun i => f i (g (some i))
```

**Theorem 13.18 (safety for arbitrary agents and an arbitrary supervisor).** Let the agents be any objects at all and let the supervisor decide in any way. In every run of the team's protocol that starts with the guardrail true, the guardrail is true of every committed state. No hypothesis is made about an agent: not about what it proposes, not about what it does in private, not about its own invariants.

```
theorem team_safe (H : Desk W → Desk W → Prop) (safe : (∀ i, W i) → Prop)
(c : ∀ i, Obj (S i)) (f : ∀ i, S i → W i)
{g0 g : ∀ o, PState (Desk W) S o}
{cfg : (teamDesign H safe c f).sys.proto.Cfg} {n : Nat}
(r : Root (teamDesign H safe c f).sys.proto g0 g cfg n)
(h0 : safe (teamProduct f g0)) : safe (teamProduct f g)
```

**Theorem 13.19.** A team of faithful, sandboxed agents meets its local obligations (Definition 12.11). The check on the orchestrator is Theorems 13.10 and 13.12; it does not mention the agents.

```
theorem teamDesign_local (H : Desk W → Desk W → Prop)
(safe : (∀ i, W i) → Prop) (c : ∀ i, Obj (S i)) (f : ∀ i, S i → W i)
(hcf : ∀ i, (c i).Faithful)
(hsand : ∀ i s s', (c i).stat s → (c i).inner s s' → f i s' = f i s) :
(teamDesign H safe c f).Local
```

**Theorem 13.20 (safety of the semantics, with agents moving by themselves).** In the nested AND an agent may also move between commits. Let the agents be faithful and let such a move leave the agent's product unchanged. Then the guardrail is true at the end of every history that starts inside the invariant.

```
theorem team_safe_sem (H : Desk W → Desk W → Prop)
(safe : (∀ i, W i) → Prop) (c : ∀ i, Obj (S i)) (f : ∀ i, S i → W i)
(hcf : ∀ i, (c i).Faithful)
(hsand : ∀ i s s', (c i).stat s → (c i).inner s s' → f i s' = f i s)
{g0 g : ∀ o, PState (Desk W) S o}
(o : ORun (teamDesign H safe c f).sys.obj g0 g)
(h0 : (teamDesign H safe c f).sys.obj.stat g0) :
safe (teamProduct f g)
```

Theorem 13.21 (the team reports advance only for a signed, agreed, safe tick). If the team's ballot is advance, then for the tentative state t it would commit: the supervisor agreed, the supervisor's record is exactly the joint product of t, the guardrail holds on it, and every agent agreed. Commits made under autonomy are child ticks rather than team ticks and are not covered by this joint-signature statement.

```
theorem team_advance (H : Desk W → Desk W → Prop) (safe : (∀ i, W i) → Prop)
(c : ∀ i, Obj (S i)) (f : ∀ i, S i → W i)
{g0 g : ∀ o, PState (Desk W) S o}
{cfg : (teamDesign H safe c f).sys.proto.Cfg} {n : Nat}
(r : Root (teamDesign H safe c f).sys.proto g0 g cfg n)
(hb : (teamDesign H safe c f).sys.proto.ballot cfg = Ballot.advance) :
H (g none) ((teamDesign H safe c f).sys.proto.tent cfg none) ∧
Desk.seen ((teamDesign H safe c f).sys.proto.tent cfg none) =
teamProduct f ((teamDesign H safe c f).sys.proto.tent cfg) ∧
safe (teamProduct f ((teamDesign H safe c f).sys.proto.tent cfg)) ∧
∀ i, (c i).vote (g (some i))
((teamDesign H safe c f).sys.proto.tent cfg (some i))
```

**Theorem 13.22 (autonomy is given only by a tick of the team).** In a run of the protocol with no tick of the team, no agent gains autonomy, and an agent that has none at the start is exactly where it started: its whole state, not only its product. Here state means the committed protocol state; tentative work and private computation are not thereby prohibited.

```
theorem team_grants (H : Desk W → Desk W → Prop) (safe : (∀ i, W i) → Prop)
(c : ∀ i, Obj (S i)) (f : ∀ i, S i → W i)
{g0 g : ∀ o, PState (Desk W) S o}
{cfg : (teamDesign H safe c f).sys.proto.Cfg} {n : Nat}
(r : Root (teamDesign H safe c f).sys.proto g0 g cfg n) (hn : n = 0) :
(∀ i, Desk.auto (g none) i = true → Desk.auto (g0 none) i = true) ∧
∀ i, Desk.auto (g0 none) i = false → g (some i) = g0 (some i)
```

**Theorem 13.23 (an agent with autonomy commits alone).** If agent i proposes t, the desk gives it autonomy and the guardrail holds on the result, the protocol commits the proposal of that agent alone. The supervisor is not asked and the team's logical time is unchanged.

```
theorem team_autonomous (H : Desk W → Desk W → Prop)
(safe : (∀ i, W i) → Prop) (c : ∀ i, Obj (S i)) (f : ∀ i, S i → W i)
{g0 g : ∀ o, PState (Desk W) S o}
{cfg : (teamDesign H safe c f).sys.proto.Cfg} {n : Nat}
(r : Root (teamDesign H safe c f).sys.proto g0 g cfg n)
(i : ι) (t : S i) (hv : (c i).vote (g (some i)) t)
(hauto : Desk.auto (g none) i = true)
(hsafe : safe (upd (teamProduct f g) i (f i t))) :
∃ cfg', Root (teamDesign H safe c f).sys.proto g0
(upd g (some i) t) cfg' n
```

Theorem 13.24 (the supervisor withdraws autonomy alone). From any point of a run, a desk change admitted by H can be realized without a team tick when it does not increase autonomy and leaves the signed record unchanged. No agent's assent is required for this path. The theorem is existential and depends on the displayed vote hypothesis; it does not guarantee withdrawal for a relation H that rejects it.

```
theorem team_stop (H : Desk W → Desk W → Prop) (safe : (∀ i, W i) → Prop)
(c : ∀ i, Obj (S i)) (f : ∀ i, S i → W i)
{g0 g : ∀ o, PState (Desk W) S o}
{cfg : (teamDesign H safe c f).sys.proto.Cfg} {n : Nat}
(r : Root (teamDesign H safe c f).sys.proto g0 g cfg n)
(d' : Desk W) (hv : H (g none) d')
(hless : ∀ i, d'.auto i = true → Desk.auto (g none) i = true)
(hseen : d'.seen = Desk.seen (g none)) :
∃ cfg', Root (teamDesign H safe c f).sys.proto g0
(upd (S := PState (Desk W) S) g none d') cfg' n
```

Theorem 13.25 (a withdrawn cycle leaves nothing behind). When the team's protocol reports retreat, which it does when every party has withdrawn, its tentative state is the committed state. Committed state can change through root commit or committed inner transitions (Definition 6.21); tentative steps and aborts do not change it.

```
theorem team_retreat (H : Desk W → Desk W → Prop) (safe : (∀ i, W i) → Prop)
(c : ∀ i, Obj (S i)) (f : ∀ i, S i → W i)
{g0 g : ∀ o, PState (Desk W) S o}
{cfg : (teamDesign H safe c f).sys.proto.Cfg} {n : Nat}
(r : Root (teamDesign H safe c f).sys.proto g0 g cfg n)
(hb : (teamDesign H safe c f).sys.proto.ballot cfg = Ballot.retreat) :
(teamDesign H safe c f).sys.proto.tent cfg = g
```

### *Teams of teams*

**Definition 13.26.** The parties of a team whose agents are themselves hierarchies.

```
def teamKidsD (H : Desk W → Desk W → Prop) (d : ∀ i, Design (S i) (W i)) :
(o : Option ι) → Design (PState (Desk W) S o) (PState (Desk W) W o)
| none => .leaf (supervisor H) (supervisor H) id
| some i => d i
```

**Definition 13.27 (a team over hierarchies).** A team whose agents are hierarchies d i, with any specification b at its root.

```
def teamOver {B : Type} (H : Desk W → Desk W → Prop)
(safe : (∀ i, W i) → Prop) (d : ∀ i, Design (S i) (W i))
(b : Obj B) (h : (∀ i, W i) → B) :
Design (∀ o, PState (Desk W) S o) B :=
Design.node (S := PState (Desk W) S) (A := PState (Desk W) W)
(teamKidsD H d) (fun g => safe (kidsOf g))
(fun _ g' => Desk.seen (g' none) = kidsOf g' ∧ safe (kidsOf g'))
(teamAllow safe) b (fun g => h (kidsOf g))
```

**Theorem 13.28 (local obligations of a team over hierarchies).** If every agent is a hierarchy that meets its local obligations and whose specification is the black box, and the team refines b, the team over them meets its local obligations. So Theorems 12.12 and 12.13 apply to teams of teams of any depth.

```
theorem teamOver_local {B : Type} (H : Desk W → Desk W → Prop)
(safe : (∀ i, W i) → Prop) (d : ∀ i, Design (S i) (W i))
(b : Obj B) (h : (∀ i, W i) → B)
(hd : ∀ i, (d i).Local) (habs : ∀ i, (d i).abs = blackbox (W i))
(hb : (Clock.team H safe).toObj.Refines b (fun g => h (kidsOf g))) :
(teamOver H safe d b h).Local
```

### *Example: two agents and a budget*

**Definition 13.29.** The products of the two agents: natural numbers.

```
– in namespace TeamExample
abbrev Wd : Bool → Type := fun _ => Nat
```

**Definition 13.30.** The states of the two agents: a pair of natural numbers.

```
– in namespace TeamExample
abbrev Sd : Bool → Type := fun _ => Nat × Nat
```

**Definition 13.31.** Two agents. The product of each is what it has spent. The guardrail: together they have spent at most 10.

```
– in namespace TeamExample
def budget (g : Bool → Nat) : Prop := g false + g true ≤ 10
```

**Definition 13.32.** A state of the team over black boxes: the grants a, and the spending w, which the desk has signed.

```
– in namespace TeamExample
def st (a : Bool → Bool) (w : Bool → Nat) : ∀ o, PState (Desk Wd) Wd o
| none => ⟨a, w⟩
| some i => w i
```

**Definition 13.33.** Agent false has spent 2, agent true nothing.

```
– in namespace TeamExample
def spent : Bool → Nat := fun b => if b then 0 else 2
```

**Definition 13.34.** Autonomy for agent true only.

```
– in namespace TeamExample
def onlyTrue : Bool → Bool := fun b => b
```

**Example 13.35.** Agent true has autonomy and 2 of 10 is spent. It may commit a spend of 3 alone.

```
– in namespace TeamExample
theorem auto_within (H : Desk Wd → Desk Wd → Prop) :
(Clock.team H budget).allow (some true) (st onlyTrue spent) (3 : Nat)
```

**Example 13.36.** With the same autonomy it may not commit a spend of 9: the orchestrator's guardrail refuses, and the supervisor is not involved.

```
– in namespace TeamExample
theorem auto_over (H : Desk Wd → Desk Wd → Prop) :
¬ (Clock.team H budget).allow (some true) (st onlyTrue spent) (9 : Nat)
```

**Example 13.37.** Agent false has no autonomy. It may commit nothing alone, whatever the amount.

```
– in namespace TeamExample
theorem no_grant (H : Desk Wd → Desk Wd → Prop) (n : Nat) :
¬ (Clock.team H budget).allow (some false) (st onlyTrue spent) n
```

**Definition 13.38.** An opaque agent. State: (spent, scratch). It proposes any state that does not un-spend. In private it may rewrite its scratch.

```
– in namespace TeamExample
def agent : Obj (Nat × Nat) :=
Obj.opaque (fun s s' => s.1 ≤ s'.1) (fun s s' => s'.1 = s.1)
```

**Definition 13.39.** The product of an agent is the first component of its state.

```
– in namespace TeamExample
def fd : ∀ i, Sd i → Wd i := fun _ s => s.1
```

**Definition 13.40.** The supervisor signs only a desk that shows a total spend of at most 5.

```
– in namespace TeamExample
def Hd : Desk Wd → Desk Wd → Prop :=
fun _ d' => d'.seen false + d'.seen true ≤ 5
```

**Definition 13.41.** The team of the two agents under supervisor Hd with guardrail budget.

```
– in namespace TeamExample
def D : Design (∀ o, PState (Desk Wd) Sd o) (∀ i, Wd i) :=
teamDesign Hd budget (fun _ => agent) fd
```

**Definition 13.42.** A state of the team: grants a, signed product seen, agent states s.

```
– in namespace TeamExample
def mk (a : Bool → Bool) (seen : Bool → Nat) (s : Bool → Nat × Nat) :
∀ o, PState (Desk Wd) Sd o
| none => ⟨a, seen⟩
| some i => s i
```

**Definition 13.43.** Nobody has autonomy, nothing spent.

```
– in namespace TeamExample
def G0 : ∀ o, PState (Desk Wd) Sd o :=
mk (fun _ => false) (fun _ => 0) (fun _ => (0, 0))
```

**Definition 13.44.** Agent false has spent 2; agent true has autonomy.

```
– in namespace TeamExample
def G1 : ∀ o, PState (Desk Wd) Sd o :=
mk onlyTrue spent (fun b => if b then (0, 7) else (2, 7))
```

**Definition 13.45.** Agent true has spent 6 on its own.

```
– in namespace TeamExample
def G2 : ∀ o, PState (Desk Wd) Sd o :=
upd (S := PState (Desk Wd) Sd) G1 (some true) ((6, 9) : Nat × Nat)
```

**Definition 13.46.** The supervisor has withdrawn all autonomy.

```
– in namespace TeamExample
def G3 : ∀ o, PState (Desk Wd) Sd o :=
upd (S := PState (Desk Wd) Sd) G2 none
(⟨fun _ => false, spent⟩ : Desk Wd)
```

**Example 13.47 (a signed tick, an autonomous commit, a withdrawal).** The protocol commits: one tick of the team, signed by the supervisor, in which agent false spends 2 and agent true is given autonomy; then a commit of agent true alone, spending 6, which the orchestrator checks against the budget and the supervisor does not see; then the supervisor alone withdraws all autonomy. The team's logical time is 1 throughout.

```
– in namespace TeamExample
theorem session : ∃ cfg', Root D.sys.proto G0 G3 cfg' 1
```

**Example 13.48.** At the end the product is 2 and 6, within the budget; the supervisor's record still shows 2 and 0, the product it signed; and it would not have signed 2 and 6.

```
– in namespace TeamExample
theorem session_end :
teamProduct fd G3 = (fun b => if b then 6 else 2) ∧
Desk.seen (G3 none) = spent ∧
∀ d d' : Desk Wd, d'.seen = teamProduct fd G3 → ¬ Hd d d'
```

### *Example: a team of teams*

**Definition 13.49.** The guardrail of an inner team: its two agents hold the same value.

```
– in namespace TeamExample
def agree (g : Bool → Nat) : Prop := g false = g true
```

**Definition 13.50.** What an inner team publishes.

```
– in namespace TeamExample
def pub (g : ∀ i, Wd i) : Nat := g false
```

**Definition 13.51.** The state of an inner team.

```
– in namespace TeamExample
abbrev InS : Type := ∀ o, PState (Desk Wd) Sd o
```

**Definition 13.52.** An inner team with supervisor Hin. Its specification is the black box (Theorem 13.14).

```
– in namespace TeamExample
def cell (Hin : Desk Wd → Desk Wd → Prop) : Design InS Nat :=
teamOver Hin agree (fun i => Design.leaf agent (blackbox Nat) (fd i))
(blackbox Nat) pub
```

**Definition 13.53.** The states of the two inner teams.

```
– in namespace TeamExample
abbrev So : Bool → Type := fun _ => InS
```

**Definition 13.54.** Two inner teams under an outer supervisor. The outer guardrail is the budget over what the inner teams publish.

```
– in namespace TeamExample
def outer (Hout : Desk Wd → Desk Wd → Prop)
(Hin : Bool → Desk Wd → Desk Wd → Prop) :
Design (∀ o, PState (Desk Wd) So o) (∀ k, Wd k) :=
teamOver Hout budget (fun k => cell (Hin k)) (guard budget) (fun g => g)
```

**Example 13.55.** The team of teams meets its local obligations, for any supervisors at both levels.

```
– in namespace TeamExample
theorem outer_local (Hout : Desk Wd → Desk Wd → Prop)
(Hin : Bool → Desk Wd → Desk Wd → Prop) : (outer Hout Hin).Local
```

**Example 13.56 (safety at depth 2).** For any supervisors at both levels, in every run from a state inside the invariant: the outer budget holds over what the inner teams publish, and in each inner team the two agents agree.

```
– in namespace TeamExample
theorem outer_safe (Hout : Desk Wd → Desk Wd → Prop)
(Hin : Bool → Desk Wd → Desk Wd → Prop)
{g0 g : ∀ o, PState (Desk Wd) So o}
{cfg : (outer Hout Hin).sys.proto.Cfg} {n : Nat}
(r : Root (outer Hout Hin).sys.proto g0 g cfg n)
(h0 : (outer Hout Hin).sys.obj.stat g0) :
budget (fun k => pub (teamProduct fd (g (some k)))) ∧
∀ k, agree (teamProduct fd (g (some k)))
```

## Axioms used by the displayed results

Of the 95 displayed items that carry a proof, 61 use no axioms. No displayed item uses Classical.choice. The others:

| Axioms | Items |
|---|---|
| Quot.sound (through function extensionality) | 6.18, 6.19, 6.27, 8.2, 8.3, 11.8, 11.9, 11.10, 11.13, 12.4, 12.6, 12.12, 12.13, 12.14, 12.15, 12.27, 12.28, 12.30, 13.10, 13.12, 13.14, 13.18, 13.19, 13.20, 13.21, 13.25, 13.28, 13.47, 13.48, 13.55, 13.56 |
| propext, Quot.sound | 7.6, 7.8, 7.9 |

## A14 Finite trace coverage and both language bounds

This section is an additional checked development, in TraceCoverage.lean. RootTrace records the initial state and each root commit, excludes committed inner moves, and embeds into the original Root relation. Its helper proofs establish that the initial state is preserved, sound protocols produce vote traces, and complete protocols realize every finite vote trace. The following displays give the definitions and resulting statements; proof terms are in the companion file.

– in namespace L0.Coverage

variable {S : Type}

Definition 14.1. Committed root traces with no inner commits.

```
inductive RootTrace (P : Proto S) (g0 : S) : Word S → P.Cfg → Nat → Prop where
  | start : RootTrace P g0 (.start g0) (P.idle g0) 0
  | step {w : Word S} {c c' : P.Cfg} {n : Nat} :
      RootTrace P g0 w c n → P.step w.last c c' → RootTrace P g0 w c' n
  | commit {w : Word S} {c : P.Cfg} {n : Nat} :
      RootTrace P g0 w c n → P.ballot c = .advance →
      RootTrace P g0 (w.extend (P.tent c)) (P.idle (P.tent c)) (n + 1)
  | abort {w : Word S} {c : P.Cfg} {n : Nat} :
      RootTrace P g0 w c n → P.ballot c = .retreat →
      RootTrace P g0 w (P.idle w.last) n
```

Definition 14.2. The finite vote-trace language from g0.

```
def Lvote (P : Proto S) (g0 : S) : Lang S :=
 fun w => w.first = g0 ∧ Trace P.obj.vote w
```

Definition 14.3. The finite root-commit language from g0.

```
def Lcommit (P : Proto S) (g0 : S) : Lang S :=
 fun w => ∃ c n, RootTrace P g0 w c n
```

Theorem 14.4. Soundness and completeness give exactly the same words.

```
theorem coverage {P : Proto S} (hs : P.Sound) (hc : P.Complete)
   (g0 : S) (w : Word S) : Lcommit P g0 w ↔ Lvote P g0 w
```

Theorem 14.5. Every state of a vote trace preserves the invariant.

```
theorem trace_safe {o : Obj S} (hf : o.Faithful) {w : Word S}
   (hw : Trace o.vote w) (h0 : o.stat w.first) : w.All o.stat
```

Theorem 14.6. Concrete required-trace witnesses give both behavioral bounds.

```
theorem contract {P : Proto S} (hs : P.Sound) (hc : P.Complete)
   (hf : P.obj.Faithful) (g0 : S) (h0 : P.obj.stat g0)
   (Lmin : Lang S) (hwitness : Lmin ⊆ Lvote P g0) :
   Lmin ⊆ Lcommit P g0 ∧ Lcommit P g0 ⊆ L_max P.obj.vote P.obj.stat
```

Definition 14.7. The image language under a fixed observation map.

```
def Observed {A : Type} (obs : S → A) (L : Lang S) : Lang A :=
 fun v => ∃ w, L w ∧ w.map obs = v
```

Theorem 14.8. Observation preserves the language correspondence.

```
theorem observed_coverage {P : Proto S} (hs : P.Sound) (hc : P.Complete)
   (g0 : S) {A : Type} (obs : S → A) (v : Word A) :
   Observed obs (Lcommit P g0) v ↔ Observed obs (Lvote P g0) v
```

Theorem 14.9. Required observable traces with witnesses are realizable.

```
theorem required_observed {P : Proto S} (hs : P.Sound) (hc : P.Complete)
  (g0 : S) {A : Type} (obs : S → A) (Lmin : Lang A)
  (hwitness : Lmin ⊆ Observed obs (Lvote P g0)) :
  Lmin ⊆ Observed obs (Lcommit P g0)
```

## Additional axiom inventory

All ten theorems in TraceCoverage.lean were checked and list no axioms. The five displayed proof-bearing items are listed here, in addition to the original 95-item appendix inventory. Required-language witnesses remain explicit hypotheses; these results do not establish fairness, deadlines, or eventual task completion.

| Item | Lean name | Axioms |
|---|---|---|
| A14.4 | L0.Coverage.coverage | none |
| A14.5 | L0.Coverage.trace_safe | none |
| A14.6 | L0.Coverage.contract | none |
| A14.8 | L0.Coverage.observed_coverage | none |
| A14.9 | L0.Coverage.required_observed | none |

## A15 Supplementary illustration of epoch witnesses

EpochBooleanNetwork.lean imports LogicalTime_NestedClocks_prototype.lean unchanged and defines L0.EpochBoolean. Two AND-gate objects retain finite histories of records. Model supplies an epoch-membership relation and three port-value functions for each gate; its time type is a parameter. A Record contains an epoch label, a witness occurrence, and the three Boolean values. Valid requires epoch membership, agreement with those port values, and the AND equation at that occurrence. Inv requires every retained record to be valid. Tick appends one valid record with the next label and preserves earlier records; the gate object votes for exactly that change and has equality as its inner relation. Junction requires the two new records' intermediate values to agree. Own requires the composed Boolean equation for every pair of valid retained records satisfying Junction. network places the gates under a parent whose joint predicate checks Junction, whose own predicate is Own, and whose allow relation permits separate child ticks. This is a historical-witness contract; it is not an input/output transfer specification for a physical circuit.

A15.1 Local obligations. gate_faithful proves the gate object faithful. numbered_tick and numbered_run prove successive record numbering from the empty history. own_of_children and network_sound establish the parent's obligations from the two child specifications; network_faithful applies Clock.faithful. empty_valid establishes the initial invariant.

A15.2 Network conclusion. compose proves Boolean substitution. parent_tick proves that every agreed parent tick from a valid state appends linked valid records, with the next epoch labels, satisfying Composed. witnessed_law relates each valid record to its gate's actual port values. physical_composition states that the second gate's witnessed output is the AND of the first gate's two inputs at its witness and the second gate's other input at its witness. The witnesses may occur at different times; the intermediate values are linked by Junction.

A15.3 Implementations and capability. implementation_verified applies Clock.verify to arbitrary faithful concrete children refining the two gate objects; it proves that the pulled clock is faithful and refines network.toObj under lift f. enabled_of_records supplies an agreed joint transition whenever the correctly numbered valid witnesses satisfy Junction. protocol_obj identifies the instantiated protocol object with network.toObj; protocol_sound and protocol_complete specialize the existing constructor theorems, using the exhaustive child list [false, true].

A15.4 Scope check. no_simultaneous_conclusion checks a two-phase counterexample: A = B = NOT t, M = C = true, and Y = t, with t Boolean. The first gate has a correct witness in phase false and the second in phase true, with matching intermediate value, yet the simultaneous three-input equation fails in both phases. This confirms that the historical conclusion cannot silently be strengthened to simultaneous combinational correctness.

A15.5 Axiom inventory. Eight of the 17 audited results list no axioms: compose, own_of_children, empty_valid, enabled_of_records, protocol_obj, witnessed_law, physical_composition, and no_simultaneous_conclusion. Six use only propext: gate_faithful, numbered_tick, numbered_run, network_sound, network_faithful, and parent_tick. protocol_sound uses only Quot.sound. implementation_verified and protocol_complete use both propext and Quot.sound. The compiler output and source checksums are retained with the proof package.

## A16 Authorized creation in an evolving forest

These declarations are from EvolvingForest.lean in Distributed_Lean_First_Checked_Step, checked with Lean 4.34.1. Proof terms remain in that file. The namespace supplies the prefix EvolvingForest to every name below; the payload type α is arbitrary. Component epochs have the meaning specified in Section 10.5.

```
namespace EvolvingForest
variable {α : Type}
```

### A16.1 Names, configurations, and local policy

A live child name extends its parent's name by an integer label. Cell stores the payload and component counters; only entries named in Config.names are committed component facts. Policy supplies the local predicates and the three instance proofs required to preserve them.

```
inductive Name where
  | root : Nat → Name
  | child : Name → Nat → Name
  deriving DecidableEq, Repr

structure Cell (α : Type) where
  value : α
  epoch : Nat
  nextChild : Nat
  birthParentEpoch : Nat

inductive Event (α : Type) where
  | genesis : Name → α → Event α
  | birth : Name → Nat → Name → α → Event α
  | tick : Name → Nat → α → Event α

structure Config (α : Type) where
  names : List Name
  cells : Name → Cell α
  ledger : List (Event α)

structure Policy (α : Type) where
  stat : Name → α → Prop
  diff : Name → α → α → Prop
  initializes : Name → α → Nat → α → Prop
  link : Name → α → Nat → α → Prop
  createAgreement : Name → α → Nat → α → Prop
  rootAgreement : Nat → α → α → Prop
  allow : Name → α → Nat → α → α → Prop
  tick_preserves : ∀ n x y, stat n x → diff n x y → stat n y
  initialize_preserves : ∀ p x k y,
    stat p x → initializes p x k y → stat (.child p k) y
  initialize_link : ∀ p x k y,
    stat p x → initializes p x k y → link p x k y
```

### A16.2 Validity and explicit state updates

Validity requires distinct live names, live parents for all children, allocation counters above every used child label, each component's static predicate, and each direct parent-child compatibility predicate. These are the precise obligations preserved by the theorem.

```
def Closed (c : Config α) : Prop :=
  ∀ p k, Name.child p k ∈ c.names → p ∈ c.names

def Sequential (c : Config α) : Prop :=
  ∀ p k, Name.child p k ∈ c.names → k < (c.cells p).nextChild

def Valid (P : Policy α) (c : Config α) : Prop :=
  c.names.Nodup ∧ Closed c ∧ Sequential c ∧
  (∀ n, n ∈ c.names → P.stat n (c.cells n).value) ∧
  (∀ p k, Name.child p k ∈ c.names →
    P.link p (c.cells p).value k (c.cells (.child p k)).value)
```

Creation uses the parent's next label, initializes the new child at epoch zero, advances the parent's epoch and allocation counter, and appends a birth event. Advancement changes one payload and its epoch, and appends a tick event. Other committed entries retain their values.

```
def newName (c : Config α) (p : Name) : Name :=
  .child p (c.cells p).nextChild

def create (c : Config α) (p : Name) (y : α) : Config α where
  names := newName c p :: c.names
  cells := fun n =>
    if n = newName c p then
      ⟨y, 0, 0, (c.cells p).epoch + 1⟩
    else if n = p then
      { c.cells n with epoch := (c.cells n).epoch + 1,
                  nextChild := (c.cells n).nextChild + 1 }
    else c.cells n
  ledger := c.ledger ++
    [.birth p ((c.cells p).epoch + 1) (newName c p) y]

def advance (c : Config α) (n : Name) (y : α) : Config α where
  names := c.names
  cells := fun m => if m = n then
    { c.cells m with value := y, epoch := (c.cells m).epoch + 1 }
    else c.cells m
  ledger := c.ledger ++ [.tick n ((c.cells n).epoch + 1) y]
```

## A16.3 Admission and committed histories

Permit requires root agreement or the immediate parent's permission. TickLinks checks compatibility for every edge incident on the changing component. The two Differential constructors specify the evidence admitting creation or advancement; neither assumes validity of the target configuration.

```
def Permit (P : Policy α) (c : Config α) (n : Name) (y : α) : Prop :=
  match n with
  | .root r => P.rootAgreement r (c.cells n).value y
  | .child p k => P.allow p (c.cells p).value k (c.cells n).value y

def TickLinks (P : Policy α) (c : Config α) (n : Name) (y : α) : Prop :=
  ∀ p k, Name.child p k ∈ c.names →
    (p = n ∨ Name.child p k = n) →
    P.link p ((advance c n y).cells p).value k
      ((advance c n y).cells (.child p k)).value

inductive Differential (P : Policy α) : Config α → Config α → Prop where
  | create (c : Config α) (p : Name) (y : α)
      (parent_exists : p ∈ c.names)
      (initialization : P.initializes p (c.cells p).value (c.cells p).nextChild y)
      (agreement : P.createAgreement p (c.cells p).value (c.cells p).nextChild y)
      (permission : Permit P c p (c.cells p).value) :
      Differential P c (EvolvingForest.create c p y)
  | tick (c : Config α) (n : Name) (y : α)
      (node_exists : n ∈ c.names)
      (law : P.diff n (c.cells n).value y)
      (permission : Permit P c n y)
      (links : TickLinks P c n y) :
      Differential P c (advance c n y)
```

CheckedRelease binds its evidence to a particular source and target. History starts at the given initial configuration and extends only through such releases; its integer index counts admitted commitments.

```
structure CheckedRelease (P : Policy α) (c : Config α) where
  target : Config α
  evidence : Differential P c target

inductive History (P : Policy α) (initial : Config α) : Config α → Nat → Prop where
  | start : History P initial initial 0
  | commit {c : Config α} {n : Nat} (h : History P initial c n)
      (r : CheckedRelease P c) : History P initial r.target (n + 1)
```

## A16.4 Three central preservation results

Theorem 16.4.1. One admitted step preserves validity. Its two cases use the policy's preservation proofs, the explicit state updates, and the required evidence for affected links.

```
theorem differential_preserves {P : Policy α} {c d : Config α}
  (hv : Valid P c) (h : Differential P c d) : Valid P d
```

Theorem 16.4.2. Every configuration reached by a finite committed history is valid when its initial configuration is valid. The proof is induction on History.

```
theorem history_preserves {P : Policy α} {initial c : Config α} {n : Nat}
  (hi : Valid P initial) (h : History P initial c n) : Valid P c
```

Theorem 16.4.3. Every finite position of a supplied unbounded run is valid. The proof is induction on the position; existence or progress of such a run is not a conclusion.

```
theorem unbounded_run_preserves (P : Policy α) (run : Nat → Config α)
  (h0 : Valid P (run 0))
  (hs : ∀ n, Differential P (run n) (run (n + 1))) :
  ∀ n, Valid P (run n)
```

## A16.5 Axiom inventory

The retained #print axioms output gives the following dependencies for the three displayed results. Definitions and transition constructors are not additional theorem counts.

| Lean declaration | Axioms |
|---|---|
| EvolvingForest.differential_preserves | propext, Quot.sound |
| EvolvingForest.history_preserves | propext, Quot.sound |
| EvolvingForest.unbounded_run_preserves | propext, Quot.sound |

The package inventory contains 45 new theorems: 13 list no axioms; 32 use only propext and/or Quot.sound. None uses Classical.choice or a custom axiom. Supporting results also check freshness, absence of parent cycles, creation epochs, namespace growth, and ledger extension. Validity here covers component predicates and direct parent-child compatibility; predicates over changing child sets require the additional obligations described in Section 10.5.